\documentclass[%
 reprint,
superscriptaddress,
 amsmath,amssymb,
 aps,
 prl,
floatfix,
]{revtex4-2}

\usepackage{graphicx}% Include figure files
\usepackage{dcolumn}% Align table columns on decimal point
\usepackage{bm}% bold math
\usepackage{hyperref}
\usepackage{xr}
\usepackage{physics}
\usepackage{chemformula}
\usepackage{siunitx}
\usepackage{pdfpages}
\usepackage{pgffor}
\makeatletter
\AtBeginDocument{\let\LS@rot\@undefined}
\makeatother

\usepackage{tikz}
\def\checkmark{\tikz\fill[scale=0.4](0,.35) -- (.25,0) -- (1,.7) -- (.25,.15) -- cycle;}

\begin{document}

\preprint{APS/123-QED}
\title{Intrinsic Fermi Surface Contribution to the Bulk Photovoltaic Effect}% Force line breaks with \\

 \author{ Lingyuan Gao}%
\thanks{These authors contributed equally}
\affiliation{Department of Chemistry,
University of Pennsylvania,
Philadelphia, PA 19104--6323, USA}
\author{Zachariah Addison}%
    \thanks{These authors contributed equally}
\affiliation{Department of Physics and Astronomy,
University of Pennsylvania,
Philadelphia, PA 19104, USA}
\author{E. J. Mele}
\affiliation{Department of Physics and Astronomy,
University of Pennsylvania,
Philadelphia, PA 19104, USA}
\author{Andrew M. Rappe}%
\affiliation{Department of Chemistry,
University of Pennsylvania,
Philadelphia, PA 19104--6323, USA}

\date{\today}% It is always \today, today,

\begin{abstract}
We study the Fermi surface contribution to the nonlinear DC photocurrent at quadratic order in a spatially uniform optical field in the ultra-clean limit.  In addition to shift and injection current, we find that polarized light incident on a metallic system generates an intrinsic contribution to the bulk photovoltaic effect deriving from photoinduced electronic transitions on the Fermi surface. In {velocity} gauge, this contribution originates in both the coherent band off-diagonal and diagonal parts of the density matrix, describing respectively, the coherent wave function evolution and the carrier dynamics of an excited population. We derive a formula for the intrinsic Fermi surface contribution for a time-reversal invariant chiral Weyl semimetal illuminated with circularly-polarized light. At low frequency, this response is proportional to the frequency of the driving field, with its sign determined by the topological charge of the Weyl nodes and with its magnitude being comparable to the recently discovered quantized circular photogalvanic effect. Our work presents a complete derivation for all contributions to nonlinear DC photocurrent and classify them according to the polarization of light in the presence and absence of TRS.
\end{abstract}

\maketitle
\noindent
{\em Introduction}---Interest in developing new platforms for efficient solar energy conversion has drawn attention to the photovoltaic properties of new materials and the physics of their light-matter interactions. The bulk photovoltaic effect (BPVE), sometimes also referred to as the photogalvanic effect (PGE), has attracted much attention, as it can directly convert light to a DC current~\cite{Sturman1992}. The BPVE is a second-order nonlinear response that can  be decomposed into terms that are  symmetric  and  antisymmetric in the polarization states of the light, corresponding to a linear photogalvanic effect (LPGE) and a circular photogalvanic effect (CPGE), respectively~\cite{Sturman1992}. %Place0:28
%The experimentally-observable BPVE has contributions from intrinsic coupling of light to electronic states, as well as the scattering of excited carriers with phonons, electrons, and other excitations.
The ballistic current is an important mechanism of the BPVE and emerges in both LPGE and CPGE \cite{alperovich1982photogalvanic,Sturman_2019,duc2019ballistic,dai2020ballistic}, %revision1
where an asymmetric distribution of photo-excited charge carriers on the conduction band are induced by electron-phonon or electron-electron scattering. %Asymmetric electron-phonon or electron-electron scattering can induce an asymmetric distribution  of photo-excited charge carriers on the conduction band which can result in a ballistic current for linear and circular polarizations of the external optical field.
Apart from the ballistic current, the CPGE is usually described as an injection current. Owing to the phase lag between orthogonal components of a circularly-polarized beam, an asymmetry in the excited state population at time-reversed momenta $\mathbf{k}$ and $-\mathbf{k}$ can be induced%, leading to a polar distribution of excited carriers in momentum space and a net current
~\cite{Sipe00p5337,laman2005ultrafast}.
In a two-band model, the carrier generation rate can be related to the Berry curvature, which relates the trace of the CPGE tensor to the quantized topological charge of degenerate points in the band structure for a Weyl semimetal~\cite{de2017quantized}.
In LPGE, other than ballistic current, shift current can be described as a coherent response associated with the real-space shift of an electron induced by a dipole-mediated vertical inter-band transition%This is described by the difference in the Berry connection between pairs of bands and the derivative of the phase of dipole matrix elements
~\cite{vonBaltz81p5590}.
Recent photo-Hall measurements of both LPGE and CPGE in an applied magnetic field have successfully separated the shift and ballistic contributions to the electric charge current \cite{burger2019direct,burger2020shift}. First principles studies ~\cite{Young12p116601, ibanez2018ab,wang2017first} of the shift current have led to the prediction and discovery of many new photovoltaic materials~\cite{tan2016enhancement, kim2017shift,zhang2018photogalvanic,Li2017RhSi,cook2017design, rangel2017large,zhang2019enhanced}.%Place1:23 
%Recently, various novel quantum materials, including topological insulators~\cite{tan2016enhancement, kim2017shift}, Weyl semimetals~\cite{zhang2018photogalvanic,Li2017RhSi}, and newly synthesized low-dimensional materials~\cite{cook2017design, rangel2017large,zhang2019enhanced} are all predicted to have large shift current responses.

Light-matter interactions in these materials are described by coupling electrons in the material to the electromagnetic potentials $A_\mu$. In {\it velocity} gauge, the external electric potential is taken to vanish, and the external electromagnetic vector potential can be incorporated into the electronic Hamiltonian through a minimal coupling procedure that augments the electronic momentum operator $\hat{\bm{p}}\rightarrow \hat{\bm{p}}+e \bm{A}(\bm{r},t)/\hbar$~\cite{passos2018nonlinear, PhysRevB.48.11705}. %Revision3
For a spatially uniform, but time varying electric field and a vanishing magnetic field, %Place2 For spatially uniform vector potentials where the electromagnetic field consists of a spatially uniform, but time varying electric field and a vanishing magnetic field,
a time-dependent gauge transformation on the electronic wave function can bring the effective Hamiltonian back to its original unperturbed form with the addition of an electric-field-induced perturbation $\delta H= e\bm{E}(t)\cdot \hat{\bm{r}}$ \cite{ventura2017gauge,rzkazewski2004equivalence}. This form of the Hamiltonian is called {\it length} gauge and has been used to derive expressions for the contributions to nonlinear electric currents in insulators and semimetals~\cite{genkin1968contribution,Sipe00p5337,Matsyshyn2019nonlinear}. %Place3:35 Here we work in {velocity} gauge, where we maintain the form of the Hamiltonian after the minimal coupling procedure has been implemented~\cite{passos2018nonlinear, PhysRevB.48.11705}.  The response to an external electric field is then formulated by identifying $\bm{E}(t)=-\partial_t \bm{A}(t)$.

Using time-dependent perturbation theory in {velocity} gauge, we derive formulas for the nonlinear photo-currents induced at quadratic order in an external driving field. %For circularly-polarized (CP) light,
We find a geometric contribution to the second-order current related to the curvature of the electronic Bloch bands that must vanish for insulating materials, but can be nonzero for metals and semimetals. This contribution to the current derives from both off-diagonal and time-independent band diagonal contributions in the density matrix, and it is proportional to $\bm{k}$-space derivatives of the unperturbed Fermi occupation factors multiplied by the band-resolved Berry curvature or quantum metric tensor \cite{provost1980riemannian, berry1984quantal}, depending on the polarization of the light.
For a time-reversal (TR) invariant system, the current is only non-vanishing under circularly-polarized (CP) illumination. By breaking time-reversal symmetry (TRS), both circularly- and linearly-polarized light can induce a current.
At zero temperature, the current derives from dipole-allowed vertical inter-band transitions for electrons whose crystal momenta are near the Fermi surface. Thus we name this new photocurrent the ``intrinsic Fermi surface contribution'', defining a new type of CPGE and LPGE response.
%Part of this intrinsic contribution is exhibited in the difference frequency response of materials~\cite{de2020difference}.

As an example, we calculate this contribution to the CPGE in a three-dimensional time-reversal symmetric, but inversion symmetry broken Weyl semimetal. We show that for an isotropic Weyl node, this response is proportional to the Weyl node's charge weighted by a factor that depends on the energy of the Bloch bands near the Fermi surface. In the small $\omega$ limit, the photovoltaic response for an isolated Weyl node is purely quantized and would therefore directly measure the topological charge of the Weyl point.
When mirror symmetries are broken in a chiral Weyl semimetal, point nodes with opposite topological charges are offset in energy, allowing a nonzero DC charge current to flow. %In the small $\omega$ limit, the photovoltaic response for an isolated Weyl node is purely quantized and would therefore directly measure the topological charge of the Weyl point.  Here, the dynamic response is an intrinsic DC charge current.
This can be distinguished from the extrinsic quantized circular photogalvanic effect deriving from the injection contribution to the current, %revision
which is proportional to a scattering time $\tau$ \cite{de2017quantized}. % and whose response describes a rate of change of the current that manifests in experiment as an extrinsic DC contribution to the current, which is proportional to a scattering time $\tau$ \cite{de2017quantized}.

\noindent
{\em General theory}---We study the Hamiltonian for a particle with charge $-e$ coupled to a time-dependent vector potential $\boldsymbol{A}(t)$
\begin{equation}
\hat{H}(\hat{\boldsymbol{r}},\hat{\boldsymbol{p}},t) = \frac{(\hat{\boldsymbol{p}}+e\boldsymbol{A}(t))^2}{2m_e} - e \ V(\hat{\boldsymbol{r}})
\end{equation}
where $V(\hat{\boldsymbol{r}})$ is the crystal potential, $m_e$ is the mass of the electron, and $\hat{\bm{p}}$ and $\hat{\bm{r}}$ are the electronic momentum and position operators.  We work in {velocity} gauge, where the response to the electric field is made by the identification $\bm{E}(t)=-\partial_t \bm{A}(t)$. Here, we focus on the single-particle Hamiltonian, $\hat{H}(\hat{r},\hat{p},t)$ and do not include the contributions deriving from many-body interactions such as excitonic effects in our model. The nonlinear DC charge current is determined by calculating the trace of the product of the velocity operator $\hat{\bm{v}}(t)=i[\hat{H}(\hat{\bm{r}},\hat{\bm{p}},t),\hat{\bm{r}}]/\hbar$ and density matrix $\hat{\rho}(t).$ To isolate the nonlinear response, we expand both $\hat{\bm{v}}(t)$ and $\hat{\rho}(t)$ in a power series of $\bm{A}(t)$ up to quadratic order in the external driving field (denoted by the superscripts on the operators below) and calculate the current as

\begin{align}
\bm{j}(t)=\dfrac{1}{V}\sum_{n=0}^{2}\text{Tr}[e\hat{\bm{v}}^{(n)}(t)\hat{\rho}^{(2-n)}(t)]
\label{curr}
\end{align}

\noindent
(See Appendix for details). 
To solve for $\hat{\rho}(t)$, we employ the von Neumann equation which describes the time evolution of this quantum operator \cite{von1927wahrscheinlichkeitstheoretischer}:
\begin{equation}
i\hbar \dfrac{d \hat{\rho}(t)}{d t}=[\hat{H}(\hat{\bm{r}},\hat{\bm{p}},t),\hat{\rho}(t)]
\label{vonN}
\end{equation}
\noindent
The Hamiltonian $\hat{H}(\hat{\bm{r}},\hat{\bm{p}},t)$ can be divided into two pieces. $\hat{H}_0(\hat{\bm{r}},\hat{\bm{p}})=\frac{\boldsymbol{p}^2}{2m_e} + V(\hat{\boldsymbol{r}})$ describes the unperturbed Hamiltonian before application of the external field and  $\hat{H}'(\hat{\bm{r}},\hat{\bm{p}},t) = \frac{e\boldsymbol{A}(t)\cdot\hat{\boldsymbol{p}}}{m_e}+\frac{e^2\bm{A}(t)\cdot \bm{A}(t)}{2m_e}$ describes the interaction between electrons in the material and the external field. 
With this substitution, we can solve equation \ref{vonN} for the density matrix order by order in the vector potential $\boldsymbol{A}(t)$. Here, for simplicity Eq.~1 starts from a non-relativistic quadratic Hamiltonian. In the Appendix, we show an equivalent derivation for a general Bloch Hamiltonian $\hat{H}(\bf{k})$.

We focus on external driving fields with few nonzero Fourier components and write $\boldsymbol{A}(t) = \sum_{\omega'}\boldsymbol{A}(\omega')e^{i(\omega'-i\eta) t}$ with $\bm{A}(\omega')=\bm{A}^*(-\omega')$.  The limit as $ \eta \rightarrow 0$ denotes an adiabatic turning on of the electromagnetic field. At second order in the electromagnetic vector potential, the current couples to two different electromagnetic fields with unique Fourier components at two general frequencies $\omega_1$ and $\omega_2$.  The DC limit is found by expanding the current in powers of $\omega_1+\omega_2$ to extract divergent and finite contributions to the current as $\omega_1\rightarrow -\omega_2$ and $\eta\rightarrow 0$.  %As discussed below, the divergent contribution to the current can be associated with the injection current, while portions of the finite contribution to the current are proportional to momentum space derivatives of the Fermi orccupation factors and vanish for insulators with no Fermi surface.

\noindent
{\em Fermi surface contribution}---We parse the nonlinear DC photocurrent into its diagonal $\boldsymbol{j}^{\rm dia}\sim\sum e\boldsymbol{v}_{nn}(\bm{k})\rho_{nn}(\bm{k})$ and  off-diagonal $\boldsymbol{j}^{\rm off}\sim \sum_{n \neq m} e\boldsymbol{v}_{nm}(\bm{k})\rho_{mn}(\bm{k})$ parts.
Here $\mathcal{O}_{nm}(\bm{k})=\bra{\Psi_n(\bm{k})}\hat{\mathcal{O}}\ket{\Psi_m(\bm{k})}$, where $\ket{\Psi_n(\bm{k})}$ are the Bloch eigenstates of the unperturbed Hamiltonian $\hat{H}_0(\hat{\bm{r}},\hat{\bm{p}})$. $\boldsymbol{j}^{\rm dia}$ describes the current generated from the dynamics of excited carrier populations: terms in $\boldsymbol{j}^{\rm dia}$ are proportional to both the velocity, $\bm{v}_{nn}(\bm{k})$, and population density, $\rho_{nn}(\bm{k})$, of Bloch electrons in bands $n$ with crystal momentum $\bm{k}$. As shown by Eqs.~(S20)-(S23) in the Appendix, $\boldsymbol{j}^{\rm dia}$ can be broken into
three pieces but belonging to two types: (I) the divergent terms $\boldsymbol{j}^{\rm dia1}$ proportional to $1/\eta\  e^{2\eta t}$ that diverge as $\eta\rightarrow 0$ and (II) $\boldsymbol{j}^{\rm dia3}$ terms that are finite as $\eta\rightarrow 0$. However, the generation rate $\partial_t \boldsymbol{j}^{\rm dia1}$ remains finite. This is called injection current $\boldsymbol{j}^{\rm inj}$, induced by resonant excitations between Bloch electrons in different bands with the same crystal momentum whose energy differs by $\hbar\omega$.

The non-divergent diagonal contribution, $\boldsymbol{j}^{\rm dia3}$, can be written as
\begin{gather}
    \boldsymbol{j}^{\rm dia3} = \frac{e^3}{2V\hbar}\sum_{n, m,i, j,\boldsymbol{k},\omega' = \pm \omega}(f^T_n(\boldsymbol{k},\mu)-f^T_m(\boldsymbol{k}, \mu))\nonumber\\\nabla_{\boldsymbol{k}}\bigg(\frac{1}{\varepsilon_{n}(\boldsymbol{k})-\varepsilon_{m}(\boldsymbol{k})+\hbar\omega'}\bigg) v_{nm}^{i}(\boldsymbol{k})v_{mn}^{j}(\boldsymbol{k})A^{i}(\omega')A^{j}(-\omega').
\end{gather}
Here, $\mu$ is the electron chemical potential. We note that  $\boldsymbol{j}^{\rm dia3}$ is rarely discussed in previous studies, but is an intrinsic contribution to the induced photocurrent: different from $\boldsymbol{j}^{\rm dia1}$ and $\boldsymbol{j}^{\rm dia2}$, it is not proportional to $1/\eta$. 

 The contribution $\boldsymbol{j}^{\rm off}$ describes a current generated from the coherence between electronic Bloch states in different bands. At second order in the perturbing field, a third band state $l$ is inserted as an intermediate transition state between the coherent pair $n$ and $m$. As shown in appendix (Eqs. (S24)(S25)), the three-band processes $n\xrightarrow{}l\xrightarrow{}m$ can be converted to an effective $n\leftrightarrow{}m$ inter-band transition with an energy factor $\frac{1}{\varepsilon_{n}(\bm{k}) - \varepsilon_{m}(\bm{k}) + \hbar\omega'-i\hbar\eta}$. This can be decomposed into a resonant $\bm{j}^{\rm off1}$ with $\delta$ function and an off-resonant $\bm{j}^{\rm off2}$ with principal part, representing two types of couplings. $\bm{j}^{\rm off1}$ is associated with the shift current $\bm{j}^{\rm shift}$, where the electron coordinate in real space is shifted along
 with the resonant excitation from band $n$ to $m$. For $\bm{j}^{\rm off2}$, after removing terms that vanish identically, it can be written as: 
\begin{align}
    \boldsymbol{j}^{\rm off2} = \frac{e^3 }{2V\hbar}\sum_{n,m,i,j,\boldsymbol{k},\omega'=\pm \omega}(f_{n}^{T}(\boldsymbol{k},\mu)-f_{m}^{T}(\boldsymbol{k},\mu))\nonumber \\\frac{1}{\varepsilon_{n}(\boldsymbol{k})-\varepsilon_{m}(\boldsymbol{k})+\hbar\omega'}\nabla_{\boldsymbol{k}}(v_{nm}^{i}(\boldsymbol{k})v_{mn}^{j}(\boldsymbol{k}))A^{i}(\omega')A^{j}(-\omega'),
\end{align}
\noindent
 We note that for insulators with a minimum band gap energy, $E_{\rm gap}$, terms contributing to both $\boldsymbol{j}^{\rm off2} $ and $\boldsymbol{j}^{\rm dia3}$ involve field-induced dipole mediated transitions between Bloch states that lead to nonzero currents even for light with frequency $|\hbar\omega|<E_{gap}$.
  
Though $\boldsymbol{j}^{\rm off2} $ and $\boldsymbol{j}^{\rm dia3}$ derive from different light-induced transition processes between Bloch states, 
the sum of the two contributions to the current can be simplified into a single term proportional to $\bm{k}$-space derivatives of the unperturbed Fermi occupation factors $f_{n}^{T}(\boldsymbol{k},\mu)$:
\begin{align}
   \boldsymbol{j}^{\rm off2}+\boldsymbol{j}^{\rm dia3}= -\sum_{n,m,i,j,\bm{k},\omega'=\pm \omega}\dfrac{e^3}{V\hbar}\bm{\nabla}_{\bm{k}}f_n^T(\bm{k},\mu)\nonumber\\ \dfrac{v_{nm}^i(\bm{k})v_{mn}^j(\bm{k})}{\varepsilon_n(\bm{k})-\varepsilon_m(\bm{k})+\hbar\omega'}A^i(\omega' )A^j(-\omega') \label{fermicurr}.
\end{align}

\noindent
 At zero temperature, $\bm{k}$-space derivatives of the Fermi occupation functions are proportional to a delta function that is nonzero only for crystal momenta along the Fermi surface.  This important simplification clarifies the absence of DC current generation for insulators perturbed by light with frequency $|\hbar \omega| < E_{\rm gap}$. In experiment, only the total current is measured, and these sub-gap currents arising from $\boldsymbol{j}^{\rm off2} $ and $\boldsymbol{j}^{\rm dia3}$ cancel, leaving only contributions to the current for materials with non-vanishing Fermi surfaces. Therefore, this contribution is nonzero only for metallic systems; %where the Fermi surface is non-vanishing;
 we denote this contribution to the current in equation \ref{fermicurr} as $\bm{j}^{\rm Fermi}$ . At finite temperature, the sharp Fermi surface will smear out, allowing states with nonzero Fermi occupations above and below the Fermi surface to contribute to generation of current. Interestingly, one recent study discovered that the sub-bandgap photocurrents can be used to probe the magnitude of finite lifetimes\cite{kaplan2020nonvanishing}, in contrast to the ``clean limit'' and independence of $\tau$ here.

For a general Bloch Hamiltonian $\hat{H}_0(\bm{k})$, in addition to $\boldsymbol{j}^{\rm Fermi}$ there is another %revision
Fermi surface contribution to the current:%that is only non-vanishing for materials with a Fermi surface:

\begin{align}
\bm{j}^{\rm Fermi2}&=-\sum_{n,i,j,\bm{k},\omega'=\pm \omega}\dfrac{e^3}{2V\hbar^3}\bra{u_n(\bm{k})}\partial_{k_i}\partial_{k_j}\hat{H}_0(\bm{k})\ket{u_n(\bm{k})} \nonumber \\
&\times\bm{\nabla}_{\bm{k}}f_n^T(\bm{k},\mu)A_i(\omega')A_j(-\omega').
\end{align}
\noindent
Derivation of this contribution is given in the general Bloch Hamiltonian part in Appendix. As shown in appendix, the sum of two Fermi surface contributions (Eq. 6 and Eq.7) is equivalent to the sum of the Drude contribution, Berry curvature dipole contribution and the free-carrier contribution derived in length gauge. The Drude and Berry curvature dipole contributions can be decomposed from $\bm{j}^{\rm Fermi2}$ but they only represent partial metallic contributions~\cite{de2020difference,watanabe2021chiral}. We note $\bm{j}^{\rm Fermi2}$ vanishes if the energy dispersion relation of $\hat{H}_0(\bm{k})$ is linear or quadratic in the crystal momentum $\bm{k}$.

By lifting TRS, $\bm{v}_{nm}(\bm{k})$ and $\bm{v}_{nm}(-\bm{k})$ are no longer negative complex conjugates. BPVE phenomena are enriched in magnetic systems, as $\bm{j}^{\rm inj}$, $\bm{j}^{\rm shift}$ and $\bm{j}^{\rm Fermi}$ are non-vanishing for driving fields with either LP and CP components. We summarize and classify our results accordingly in Table \ref{table:1}.
{\color{red}{
\begin{table*}
\begin{tabular}{ |p{2.2cm}|p{1.5cm}|p{1.5cm}|p{1.8cm}|p{1.8cm}|p{1cm}|p{1cm}|p{1cm}|p{1cm}|}
 \hline
 {\bf Contribution}& \multicolumn{2}{|c|}{{\bf Type}} & \multicolumn{2}{|c|}{{\bf Origin}} & \multicolumn{2}{|c|}{\bf{TR invariant}} &
 \multicolumn{2}{|c|}{\bf{TR broken}}\\
 \hline
  & Intrinsic & Extrinsic & Diagonal & Off-diagonal & LP  & CP &LP & CP\\
 \hline
 Injection & & \checkmark & \checkmark &  & & \checkmark & \checkmark & \checkmark\\
 \hline
 Ballistic & & \checkmark & \checkmark &  & \checkmark & \checkmark & \checkmark& \checkmark\\
 \hline
 Shift & \checkmark & &  & \checkmark & \checkmark &  &\checkmark & \checkmark \\ 
 \hline
 Fermi surface & \checkmark & & \checkmark & \checkmark & & \checkmark & \checkmark & \checkmark \\
 \hline
 Fermi surface2 & \checkmark & & \checkmark & \checkmark & \checkmark &  & \checkmark & \\
 \hline
\end{tabular}
 \caption{The classification of all types of contributions to the nonlinear DC photocurrent.  ``Intrinsic/extrinsic'' means whether the current is independent/dependent on $\tau$.  ``Diagonal/off-diagonal'' refers to the part of $\rho$ from which the current arises. The last four columns give the correspondence between each current and light polarization with and without TRS. Each contribution is gauge invariant, independent of the choice of phase of the Bloch wavefunctions.}
 \label{table:1}
\end{table*}

}}
\noindent
{\em Example: single Weyl point}---We calculate this new Fermi surface contribution $\bm{j}^{\rm Fermi}$ for a single isotropic Weyl node in a minimal two-band model:
\begin{align}
    H_{0} = v_{0}\sum_{\alpha = x,y,z}k_{\alpha}\sigma_{\alpha} 
\end{align}
  Here $v_{0}$ is the Fermi velocity, and $\sigma_{\alpha}$ are the Pauli matrices. The system is illuminated by CP light described by the vector field $\bm{A}(t)=A_0(\cos(\omega t)\hat{\bm{x}}+\sin(\omega t)\hat{\bm{y}})$ that propagates in the $\hat{\bm{z}}$-direction. With linearly dispersive bands, $\bm{j}^{\rm Fermi2}$ is zero.
  
At zero temperature, we write $\boldsymbol{j}^{\rm Fermi}$ in a different form:
\begin{align}
    \boldsymbol{j}^{\rm Fermi}=\frac{e^3\omega i}{V\hbar^2}\sum_{n,m,i,j,\bm{k}}\dfrac{(\varepsilon_n(\bm{k})-\varepsilon_m(\bm{k}))^2 }{(\varepsilon_n(\bm{k})-\varepsilon_m(\bm{k}))^2-\hbar^2\omega^2} \nonumber \\
\times \delta(\varepsilon_{n}(\bm{k})-\mu)\bm{v}_{nn}(\bm{k})\tilde{\Omega}_{nm}^{ij}(\bm{k})A^{i}(\omega)A^{j}(-\omega),
\end{align}
where $\tilde{\Omega}^{ij}_{nm}(\bm{k})=-i(R^i_{nm}(\bm{k})R^j_{mn}(\bm{k})-R^j_{nm}(\bm{k})R^i_{mn}(\bm{k}))$ is the band-resolved Berry curvature. In this way, the DC charge current can be related with the integration of a weighted Berry curvature dipole over the Brillouin zone~\cite{Fu2015nonlinear,rostami2018nonlinear, Matsyshyn2019nonlinear}. With $\bm{A}(t)$ introduced above, $\bm{j}^{\rm Fermi}$ only has a $\hat{\bm{z}}$ component,
 \begin{align}
     {j}^{{\rm {Fermi}}, z}= \frac{e^3 i }{V\hbar^{2}}\frac{\omega\boldsymbol(2\mu)^{2}}{(2\mu)^{2}-\hbar^{2}\omega^{2}}\sum_{\boldsymbol{k}}\delta(k-k_F)\nonumber\\\times\hat{n}^{z}(\boldsymbol{k})\Omega^{z}(\boldsymbol{k})A^{x}(\omega)A^{y}(-\omega),
 \end{align}
where $\hat{n}^{z}(\boldsymbol{k})$ is the unit vector normal to the Fermi surface, $\Omega^z(\bm{k})$ is the $\hat{\bm{z}}$-component of the Berry curvature, and $\mu$ is the chemical potential. For this simple model, $|\varepsilon_n(\bm{k})-\varepsilon_m(\bm{k})|=2|\mu|$ and is constant across the Fermi surface.  For arbitrary CP light illumination, the current can be written as $j^{{\rm Fermi},i} =\sum_j \chi_{ij}(\omega)[\boldsymbol{E}(\omega)\times\boldsymbol{E}^{*}(\omega)]_{j}$, where $\chi_{ij}(\omega)$ is a purely imaginary photovoltaic tensor with the property

 \begin{align}
     \text{Tr}[\chi_{ij}(\omega)]=i\frac{e^3 }{h^{2}}\frac{(2\mu)^{2}}{\omega[(2\mu)^{2}-\hbar^{2}\omega^{2}]}Q_{n},
 \end{align}
 where $Q=\frac{1}{2\pi}\int_{FS} d\boldsymbol{S}\cdot\boldsymbol{\Omega}(\boldsymbol{k})$ is the charge of the Weyl point. In the limit of $|\hbar \omega|\ll |\mu|$, the current is proportional to the Weyl node's charge, and the trace of $\chi(\omega)$ is proportional to the Chern number of the Fermi surface which encloses this topological degeneracy.
The charge is a purely topological aspect of the band structure and is robust even when the isotropic symmetry of the Weyl node is broken and the Fermi surface degenerates into an elliptic surface. Consistent with the injection current, the sign of the photocurrent is also dictated by the charge of the Weyl point, and can be used to detect the chirality of the band singularity~\cite{ma2017direct}. We note that for a tilted Weyl cone the energy difference between bands is not constant along the Fermi surface and the $\text{Tr}[\chi_{ij}(\omega)]$ cannot be reduced to the simple form given in equation 11.

 Unlike the quantized circular photogalvanic effect  \cite{de2017quantized}, here the current does not originate from coherence of optically coupled band states, but instead originates from electrons with crystal momentum along the Fermi surface. In addition, the effect is intrinsic and only depends on the light frequency $\omega$ and not on an  extrinsic scattering time $\tau$.
 This distinguishes itself from the injection current in that due to its coherent nature it does not require a finite $\tau$ to populate carriers~\cite{ma2019nonlinear}. For metallic systems illuminated by mid-infrared light, $\bm{j}^{\rm Fermi}$ is significant and comparable to $\bm{j}^{\rm inj}$. Unlike the difference frequency generation scenario studied in other work~\cite{de2020difference}, $\bm{j}^{\rm Fermi}$ can be excited by one monochromatic, polarized light.

\begin{figure*}
\includegraphics[width = 160mm]{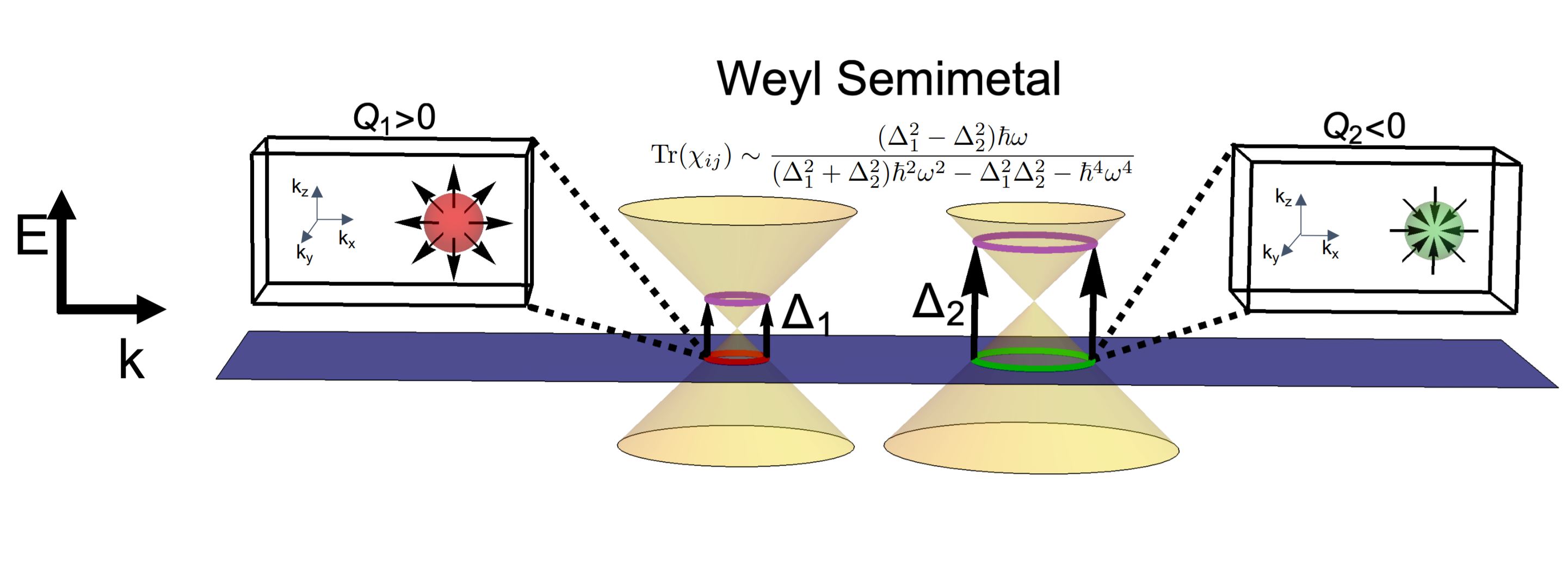}
\caption{\label{fig1} Illustration of two Weyl points with opposite charge $Q_n$ near the Fermi surface.  The Fermi surface is shown by the blue plane that intersects Weyl cones at the red and green conic sections. Optical excitations are along the Fermi surface and transition energies  ($\Delta_1$ and $\Delta_2$) are different at two cones.}
\end{figure*}

\begin{figure}
\includegraphics[width = 80mm]{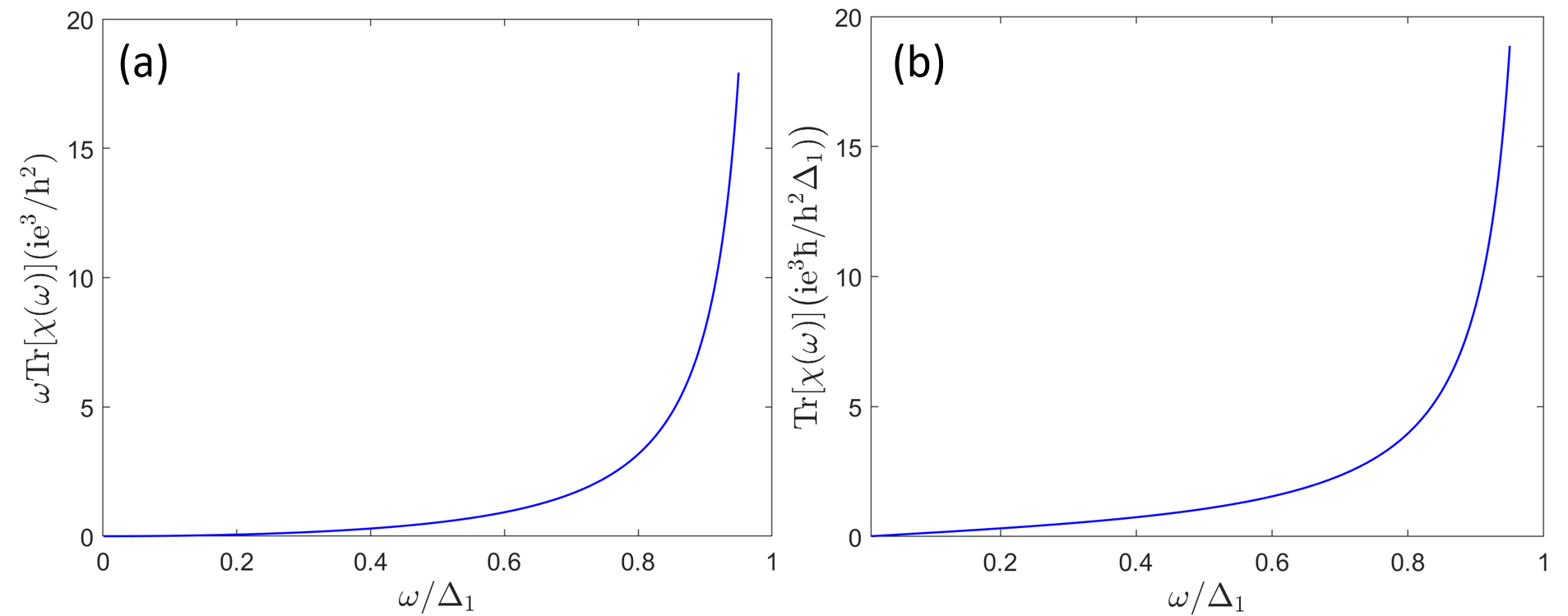}
\caption{\label{fig2} The relation between  $\omega$ and (a) $\omega \text{Tr}[\chi(\omega)]$ (b) $\text{Tr}[\chi(\omega)]$ for a four Weyl nodes system with $Q_{1} = Q_{2} = 1$, $\Delta_{1} = \Delta_{2}$ and $Q_{3} = Q_{4} = -1$, $\Delta_{3} = \Delta_{4} = 2\Delta_{1}$.  }
\end{figure}

\noindent
 {\em Example: multiple Weyl nodes}---
 We now demonstrate calculation of $\bm{j}^{\rm Fermi}$ for a time-reversal symmetric, but inversion and mirror broken Weyl semimetal. For systems with TRS, Weyl points in the band structure must come in pairs.  The two Weyl points in a pair have the same energy and topological charge, but are located at points in the Brillouin zone with opposite crystal momentum.  The topology of the Brillouin zone demands that the sum of the charges of all Weyl points in the Brillouin zone must vanish: $\sum \limits_n Q_n=0$ forcing the number of Weyl points in a time-reversal symmetric system to be a multiple of four.  If mirror symmetry is broken, Weyl points of opposite sign need not occur at the same energy~\cite{de2017quantized}, allowing the energy differences between bands, $\Delta(\bm{k})=\epsilon_n(\bm{k})-\epsilon_m(\bm{k})$, near each Weyl point to be inequivalent (see Figure \ref{fig1}).  We have seen that for a single Weyl point and for light illumination $|\hbar\omega|\ll|\mu|$, the trace of $\chi(\omega)$ is simply proportional to the integral of the Berry curvature across the Fermi surface. For a system of isotropic Weyl points illuminated by light with arbitrary $\omega$, $\Delta(\bm{k})\rightarrow \Delta$ and we may write the trace of $\chi(\omega)$ as 
\begin{align}
     \text{Tr}[\chi_{ij}(\omega)]= i\frac{e^3 }{h^2}\sum_{n}\frac{(\Delta_{n})^{2}}{\omega[(\Delta_{n})^{2}-\hbar^{2}\omega^{2}]}Q_{n},
\end{align}
Here $Q_n$ is the charge of node $n$ and $\Delta_n$ is the energy difference between Bloch states with crystal momentum along the Fermi surface near Weyl node $n$. With mirror symmetry, $\Delta_{n}$ are the same for differently-charged Weyl points. No charge current flows, but only a chiral current represents a charge pumping between Weyl nodes with different chiralities, which is not observable and is similar to the chiral anomaly~\cite{zyuzin2012topological,burkov2015chiral,burkov2018weyl}. If $\Delta_{n}$ are different, the trace of $\chi(\omega)$ will be nonzero. 
Unlike the quantized CPGE induced by injection current where Pauli blocking can forbid the photocurrent when upper state is filled \cite{de2017quantized}, the chemical potential can sit either above or below the Weyl node. Each node's contribution to the current does not change its sign whether the Fermi surface is an electron or hole pocket.  In the limit $|\hbar\omega| \ll |\Delta_{n}|$, we can expand Tr[$\chi_{ij}(\omega)]$ in powers of $\hbar^2\omega^2/(\Delta_n)^2$,

 \begin{align}
     \text{Tr}[\chi_{ij}(\omega)]&\approx i\frac{e^3 }{\omega h^2}\sum_{n}\bigg(1+\bigg(\frac{\hbar\omega}{\Delta_{n}}\bigg)^2+\mathcal{O}\bigg(\dfrac{\hbar \omega}{\Delta_n}\bigg)^4\bigg)Q_{n}\nonumber\\
     &\approx\frac{ie^3\omega}{(2\pi)^{2}}\sum_{n}\dfrac{Q_n}{(\Delta_{n})^2} .
\end{align}
We see that the leading-order term in the expansion is linear in $\omega$, with slope determined by the ratio of the charges of Weyl nodes to the energy differences of the Bloch bands near the Weyl nodes.   The relationships between $\omega$ and  $\omega\text{Tr}[\chi(\omega)]$, and $\omega$ and $ \text{Tr}[\chi(\omega)]$ for a minimal four-Weyl node system are plotted in Fig.\ 2.  Breaking the isotropy of the Weyl nodes takes $\Delta_n\rightarrow \Delta_n(\bm{k})$, and the leading contribution to the trace of $\chi(\omega)$ will no longer be directly proportional to the charges of the Weyl nodes. However, the first-order non-vanishing contribution to the current will maintain a linear relationship to the frequency of light. 

%\noindent
\noindent
 {\em Candidate materials}---Noncentrosymmetric metals characterized by the coexistence of metallicity and ferroelectric distortions provide a promising platform for observing this novel addition to the nonlinear current.  Experiments have demonstrated that metallic \ch{LiOsO3} and \ch{Cd2Re2O7}  experience a centrosymmetric to non-centrosymmetric phase transition at 140~K and at 200~K, respectively~\cite{sergienko2004metallic,shi2013ferroelectric}; while the engineering of interfaces in \ch{ANiO3}/\ch{LaAlO3} heterostructures provides another scheme for achieving other interesting noncentrosymmetric metals~\cite{kim2016polar}. In addition, recent studies on few-layer topological semimetal \ch{WTe2} have demonstrated a switchable ferroelectric polarization that could also provide a platform to observe large $\boldsymbol{j}^{\rm Fermi}$ under  illumination by CP light~\cite{fei2018ferroelectric}.    

\noindent{\em Acknowledgement}---L.G. and A.M.R. acknowledge the support of the US Department of Energy, Office of Basic Energy Sciences, under grant number DE-FG02-07ER46431. ZA and EJM are supported by the Department of Energy under Grant No. DE-FG02-84ER45118. L. G. led the analytic derivations and performed all of the numerical calculations. Z. A. participated actively in the derivations and their interpretation. L. G. and Z. A. wrote the manuscript, and E. J. M. and A. M. R. edited the manuscript. E. J. M. and A. M. R. supervised all aspects of the project. We acknowledge F.de Juan, T.Morimoto, J.E.Moore and A.G.Grushin for sharing their work with us and point out a similar form of equation \ref{fermicurr} can also be derived in length gauge \cite{de2020difference}. We also acknowledge Hiruku Watanabe and Yoichi Yanese for informing us of the update of ~\cite{watanabe2021chiral}. We finally acknowledge Daniel Kaplan for useful discussion at the finite lifetime limit~\cite{kaplan2020nonvanishing}.

\bibliography{ref.bib}% Produces the bibliography via BibTeX.

\foreach \x in {1,...,16}
{
\clearpage
\includepdf[pages={\x,{}}]{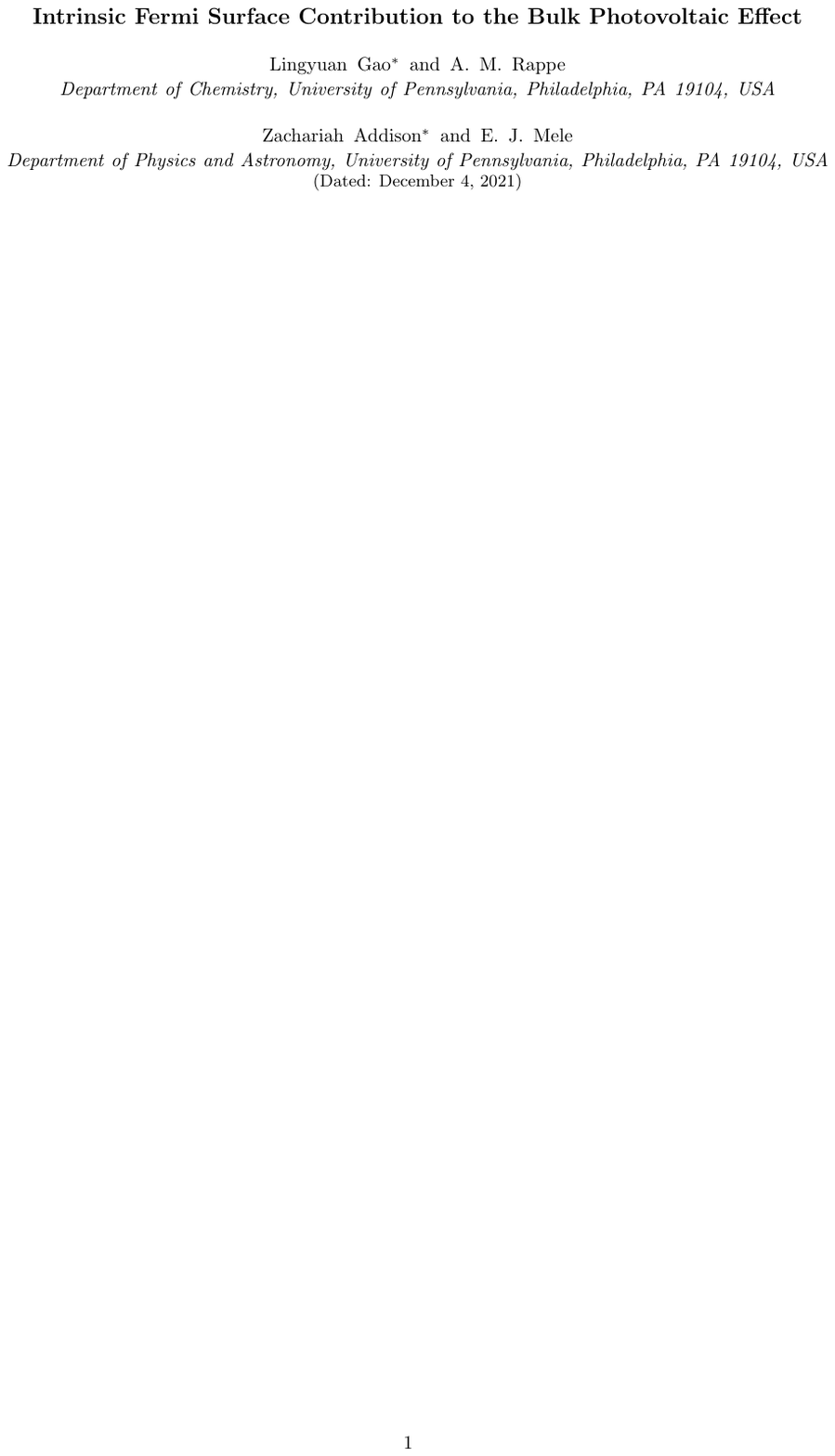}
}

\end{document}

\preprint{}

\title{Intrinsic Fermi Surface Contribution to the Bulk Photovoltaic Effect}% Force line breaks with \\

 \author{ Lingyuan Gao}%
\thanks{These authors contributed equally}
\affiliation{Department of Chemistry,
University of Pennsylvania,
Philadelphia, PA 19104, USA}
\author{Zachariah Addison}%
    \thanks{These authors contributed equally}
\affiliation{Department of Physics and Astronomy,
University of Pennsylvania,
Philadelphia, PA 19104, USA}
\author{E. J. Mele}
\affiliation{Department of Physics and Astronomy,
University of Pennsylvania,
Philadelphia, PA 19104, USA}
\author{A. M. Rappe}%
\affiliation{Department of Chemistry,
University of Pennsylvania,
Philadelphia, PA 19104, USA}

\date{\today}% It is always \today, today,

\maketitle

\section{Nonlinear DC photocurrent: Infinite Bands Limit}
We start from a quadratic Hamiltonian 
\begin{equation}
H_{0}(\boldsymbol{r},\boldsymbol{p}) = \frac{\boldsymbol{p}^2}{2m_e} + V(\boldsymbol{r}).  
\end{equation}
 At infinite bands limit, the commutation relation holds\cite{ventura2017gauge,passos2018nonlinear}: 
\begin{equation}
    [{r}^{i},{p}^{j}] = i\hbar\delta^{ij}.
\end{equation}
We define the velocity operator as
\begin{equation}
    \boldsymbol{\hat{v}} = \frac{i}{\hbar}[\hat{H}(\hat{\bm{r}},\hat{\bm{p}},t),\hat{\boldsymbol{r}}].
\end{equation}
 For Hamiltonian (S1), $\boldsymbol{\hat{v}} = \frac{\hat{\boldsymbol{p}}}{m_e}$. Taking Bloch state $|\chi _n(\bm{k})\rangle$ as basis, velocity matrix elements $\boldsymbol{v}_{nm}(\boldsymbol{k})$ are diagonal of $\boldsymbol{k}$: 
\begin{equation}
    \boldsymbol{v}_{nm}^{\boldsymbol{k},\boldsymbol{k'}}= (2\pi)^{d}\delta(\boldsymbol{k}-\boldsymbol{k'})\boldsymbol{v}_{nm}(\boldsymbol{k}),
\end{equation}
and can be computed as
\begin{align}
    \boldsymbol{v}_{nm}(\boldsymbol{k})=\frac{\hbar}{m_e}\langle \chi_n(\boldsymbol{k})|-i\nabla|\chi_m(\boldsymbol{k})\rangle =\frac{\hbar}{m_e}\langle u_n(\boldsymbol{k})|-i\nabla+\boldsymbol{k}|u_m(\boldsymbol{k})\rangle,
\end{align}
where $u_{n}(\boldsymbol{k})$, $u_{m}(\boldsymbol{k})$ are corresponding periodic functions of $\chi_{n}(\boldsymbol{k})$ and $\chi_{m}(\boldsymbol{k})$, respectively. When $n \neq m$, $\bm{v}_{nm}(\boldsymbol{k})$ can be replaced with the position matrix element $\bm{r}_{nm}(\boldsymbol{k})$: 
\begin{equation}
\boldsymbol{v}_{nm}(\boldsymbol{k}) = \frac{i(\varepsilon_{n}(\boldsymbol{k}) - \varepsilon_{m}(\boldsymbol{k}))}{\hbar} \boldsymbol{r}_{nm}(k),
\label{vtor}
\end{equation}
and $\bm{r}_{nm}(\boldsymbol{k})$ is defined through:
\begin{align}
    \boldsymbol{r}_{nm}^{\boldsymbol{k,k'}}=\langle\chi_{n}(\boldsymbol{k})|\vec{r}|\chi_{m}(\boldsymbol{k'})\rangle=-i\delta_{nm}(2\pi)^{d}\nabla_{\boldsymbol{k}} \delta({\boldsymbol{k}-\boldsymbol{k'}}) \nonumber + \delta({\boldsymbol{k}-\boldsymbol{k'}})(2\pi)^{d}\langle u_{n}(\boldsymbol{k'})|i\nabla_{\boldsymbol{k}} |u_{m}(\boldsymbol{k})\rangle\nonumber\\
    = -i\delta_{nm}(2\pi)^{d}\nabla_{\boldsymbol{k}} \delta({\boldsymbol{k}-\boldsymbol{k'}}) \nonumber + \delta({\boldsymbol{k}-\boldsymbol{k'}})(2\pi)^{d}\bm{r}_{nm}(\bm{k}).
\end{align}
The first term contains a $\boldsymbol{k}$ derivative to the $\delta$ function, which is not well defined mathematically. It vanishes unless $n = m$. For $n = m$, we integrate it over $ k = k'$:
\begin{equation}
\int_{\mathbf{k}-\epsilon}^{{\mathbf{k}+\epsilon}}d^{3}\mathbf{k'} g(\mathbf{k'})    \boldsymbol{r}_{nn}^{\boldsymbol{k,k'}}=i\nabla_{\mathbf{k}} g(\mathbf{k}) + g(\mathbf{k})\boldsymbol{R}_{nn}(\boldsymbol{k}),
\end{equation}
\begin{equation}
\int_{\mathbf{k}-\epsilon}^{{\mathbf{k}+\epsilon}}d^{3}\mathbf{k'} g(\mathbf{k'})    \boldsymbol{r}_{nn}^{\boldsymbol{k',k}}=-i\nabla_{\mathbf{k}} g(\mathbf{k}) + g(\mathbf{k})\boldsymbol{R}_{nn}(\boldsymbol{k}),
\end{equation}
where $\epsilon \xrightarrow{} 0$. In the equation above we replace $\bm{r}$ with $\bm{R}$ to denote the band-resolved Berry connection. Below we show explicitly how to use commutation relation and position matrix elements to derive nonlinear DC photocurrent at infinite bands limit.

Under velocity gauge, through minimal coupling, the Hamiltonian becomes:
\begin{equation}
H(\boldsymbol{r},\boldsymbol{p},t) = \frac{(\boldsymbol{p}+e\bm{A}(t))^2}{2m_e} + V(\boldsymbol{r}).  
\end{equation}
From (S3), the velocity operator becomes:
\begin{equation}
    \boldsymbol{\hat{v}}^{A} = \frac{\hat{\boldsymbol{p}} + e\bm{A}(t)}{m_e},
\end{equation}
and the corrected velocity matrix elements $\bm{v}^{A}_{nm\bm{k}}(t)$ is
\begin{equation}
    \bm{v}^{A}_{nm\bm{k}}(t) = \bm{v}_{nm}(\boldsymbol{k})+\delta_{nm}\frac{e}{m_e}\bm{A}(t),
\end{equation}
where the superscript ``$A$'' denotes the minimal coupling. We note in (S11), the correction to velocity matrix element is only at the first order of vector potential $\bm{A}$: all higher order corrections vanish due to commutation relation (S2) \cite{ventura2017gauge}.

To describe the interaction between the external field and electrons of the system, the perturbation is written as
\begin{equation}
    H'(t) = \frac{e}{m}\vec{P}\cdot\vec{A}(t)e^{\eta t}
\end{equation}
where $\eta \xrightarrow{} 0^{+}$ denotes the perturbation is turned on adiabatically at $t \xrightarrow{} - \infty$. The $\frac{e^2 {A}^{2}(t)}{2m_e}$ is dropped as it is proportional to the identity and will not contribute to the current at quadratic order in the electromagnetic field. For time-dependent perturbation, Von Neumann equation describes the time evolution of the density matrix $\rho (t)$:
\begin{equation}
    \frac{\partial{\rho_{I}}}{\partial t} = -\frac{i}{\hbar}\Big[H'_{I}(t),\rho_{I}(t)\Big],
\end{equation}
where subscript ``${I}$'' denotes the interaction picture.
By integrating equation (S13), we can solve $\rho_{I} (t)$ order by order:
\begin{multline}
\rho_{I}(t) = \rho_{I}^{(0)}(t)+\rho_{I}^{(1)}(t)+\rho_{I}^{(2)}(t)+...
\\ = \rho_{I}(-\infty)-\frac{i}{\hbar}\int_{-\infty}^{t} dt'\Big[H'_{I}(t'),\rho_{I}(-\infty)] + (-\frac{i}{\hbar})^{2}\int_{-\infty}^{t}dt'\int_{-\infty}^{t'}dt''[H'_{I}(t'),[H'_{I}(t''),\rho_{I}(-\infty)]]+...
\end{multline}
Here we only present the first order $\rho^{(1)}(t)$ and the second order $\rho^{(2)}(t)$:
\begin{equation}
    \rho^{(1)}_{nm\bm{k}}(t) = e\sum_{i,\omega_1 =\pm \omega, }\frac{\big(f_{n}^{T}(\boldsymbol{k},\mu)-f_{m}^{T}(\boldsymbol{k},\mu)\big)v_{nm}^{i}(\bm{k})A_{i}(\omega_{1})}{\varepsilon_{n} - \varepsilon_{m} +\hbar \omega_{1} - i\hbar \eta }e^{i\omega_1 t}e^{\eta t},
\end{equation}
\begin{widetext}
\begin{align}
    \rho^{(2)}_{nm\bm{k}}(t) = e^{2}\sum_{i,j, l,\omega_1 =\pm \omega, \omega_2=\pm \omega}\Bigg\{ \frac{(f_{m}^{T}(\boldsymbol{k},\mu)-f_{l}^{T}(\boldsymbol{k},\mu))v^{j}_{nl}(\boldsymbol{k})v^{i}_{lm}(\boldsymbol{k})A_{i}(\omega_1)A_{j}(\omega_2)}{(\varepsilon_{l}(\boldsymbol{k})-\varepsilon_{m}(\boldsymbol{k})+\hbar\omega_1-i\hbar\eta)(\varepsilon_{n}(\boldsymbol{k})-\varepsilon_{m}(\boldsymbol{k})+\hbar(\omega_{1}+\omega_{2})-2i\hbar\eta)} \nonumber\\+ 
    \frac{(f_{n}^{T}(\boldsymbol{k},\mu)-f_{l}^{T}(\boldsymbol{k},\mu))v^{i}_{nl}(\boldsymbol{k})v^{j}_{lm}(\boldsymbol{k})A_{i}(\omega_1)A_{j}(\omega_2)}{(\varepsilon_{n}(\boldsymbol{k})-\varepsilon_{l}(\boldsymbol{k})+\hbar\omega_1-i\hbar\eta)(\varepsilon_{n}(\boldsymbol{k})-\varepsilon_{m}(\boldsymbol{k})+\hbar(\omega_{1}+\omega_{2})-2i\hbar\eta)}\Bigg\}e^{i (\omega_1 +\omega_2) t} e^{2\eta t}. 
\end{align}
\end{widetext}
The macroscopic current is computed as:
\begin{equation}
    \bm{j}(t) = \frac{e}{V}\text{Tr}[\bm{v}(t)\rho(t)].
\end{equation}
By separating $\bm{v}(t)$ and $\rho(t)$ into different orders of $\bm{A}$ according to (S11) and (S14), the nonlinear photocurrent is:
\begin{equation}
    \bm{j}(t) = \frac{e}{V}\sum_{n,m,\bm{k}}[\bm{v}_{mn}(\bm{k})\rho^{(2)}_{nm}(t)] + \frac{e}{V}\sum_{n,m,\bm{k}}[\delta_{mn}\frac{e}{m_e}\bm{A}\rho^{(1)}_{nm\bm{k}}(t)].
\end{equation}
Since $\bm{v}^{A}_{mn\bm{k}}(t)$ only has the first order correction of $\bm{A}$ on diagonal terms, while according to (S15) the correction on $\rho^{(1)}_{nm\bm{k}}(t)$ is only on off-diagonal terms, the nonlinear photocurrent only comes from $v_{mn}(\bm{k})\rho^{(2)}_{nm\bm{k}}(t)$. In (S16) we require $\omega_1 + \omega_2 = 0$ as the condition for DC response, thus $\rho^{2}_{nm\bm{k}}(t)$ is reduced to:
\begin{widetext}
\begin{align}
    \rho^{(2)}_{nm\bm{k}}(t) = e^{2}\sum_{i,j, l,\omega' =\pm \omega} \frac{(f_{m}^{T}(\boldsymbol{k},\mu)-f_{l}^{T}(\boldsymbol{k},\mu))v^{j}_{nl}(\boldsymbol{k})v^{i}_{lm}(\boldsymbol{k})A_{i}(\omega')A_{j}(-\omega')}{(\varepsilon_{l}(\boldsymbol{k})-\varepsilon_{m}(\boldsymbol{k})+\hbar\omega'-i\hbar\eta)(\varepsilon_{n}(\boldsymbol{k})-\varepsilon_{m}(\boldsymbol{k})-2i\hbar\eta)} \nonumber\\+ 
    \frac{(f_{n}^{T}(\boldsymbol{k},\mu)-f_{l}^{T}(\boldsymbol{k},\mu))v^{i}_{nl}(\boldsymbol{k})v^{j}_{lm}(\boldsymbol{k})A_{i}(\omega')A_{j}(-\omega')}{(\varepsilon_{n}(\boldsymbol{k})-\varepsilon_{l}(\boldsymbol{k})+\hbar\omega'-i\hbar\eta)(\varepsilon_{n}(\boldsymbol{k})-\varepsilon_{m}(\boldsymbol{k})-2i\hbar\eta)}e^{2\eta t}. 
\end{align}
\end{widetext}

As described in the main manuscript, we separate $\boldsymbol{j}({t})$ into diagonal contribution $\boldsymbol{j}^{\rm dia} =  \frac{1}{V}\sum_{n,\bm{k}}\bm{v}_{nn}(\bm{k})\rho^{2}_{nn}(t)$ and off-diagonal contribution $\boldsymbol{j}^{\rm off} =  \frac{1}{V}\sum_{n,m,n\neq m, \bm{k}}\bm{v}_{mn}(\bm{k})\rho^{2}_{nm}(t)$.  First we deal with $\bm{j}^{\rm dia}$. Using identity:

\begin{align}
\lim_{\eta\rightarrow0}\dfrac{1}{-2i\hbar\eta}\dfrac{1}{\varepsilon_n(\bm{k})-\varepsilon_m(\bm{k})+\hbar\omega-i\hbar\eta}=\lim_{\eta\rightarrow0}\bigg(-\dfrac{\pi}{2\hbar\eta}\delta(\varepsilon_n(\bm{k})-\varepsilon_m(\bm{k})
+\hbar\omega) \nonumber\\
-\dfrac{1}{2i\hbar\eta}\mathcal{P}\dfrac{1}{\varepsilon_n(\bm{k})-\varepsilon_m(\bm{k})+\hbar\omega}-\dfrac{1}{2}\mathcal{P}\dfrac{1}{(\varepsilon_n(\bm{k})-\varepsilon_m(\bm{k})+\hbar\omega)^2}\bigg),
\label{dia_factor}
\end{align}

$\boldsymbol{j}^{\rm dia}$ can be broken into three pieces. The contribution from the first resonant term is: 
\begin{widetext}
\begin{align}
    \boldsymbol{j}^{\rm dia1} =-\frac{e^3\pi }{2 V\hbar\eta}\sum_{n,m,i,j,\omega',\boldsymbol{k}} (f_{n}^{T}(\boldsymbol{k},\mu)-f_{m}^{T}(\boldsymbol{k},\mu))(\boldsymbol{v}_{nn}(\boldsymbol{k})-\boldsymbol{v}_{mm}(\boldsymbol{k}))v_{nm}^{i}(\boldsymbol{k})v_{mn}^{j}(\boldsymbol{k})
    \delta(\varepsilon_{n}-\varepsilon_{m}+\hbar\omega')A^{i}(\omega')A^{j}(-\omega').
\label{j_dia1}
\end{align}
\end{widetext}
This is generally referred to as injection current ($\bm{j}^{\rm inj}$). For systems with time-reversal symmetry (TRS), $\boldsymbol{v}_{nn}(\boldsymbol{k}) = -\boldsymbol{v}_{nn}(-\boldsymbol{k})$, injection current is non-vanishing only with CP light. The second principal term vanishes trivially by switching indices $i\leftrightarrow j,n\leftrightarrow m, \omega'\leftrightarrow -\omega'$. For the third term, using the identity
\begin{equation}
-\mathcal{P}\dfrac{\bm{v}_{nn}(\bm{k})-\bm{v}_{mm}(\bm{k})}{2(\varepsilon_n(\bm{k})-\varepsilon_m(\bm{k})+\hbar\omega)^2}=\bm{\nabla}_{\bm{k}}\mathcal{P}\dfrac{1}{2\hbar(\varepsilon_n(\bm{k})-\varepsilon_m(\bm{k})+\hbar\omega)},
\label{group-derivative}
\end{equation}
it can be simplified as:
\begin{widetext}
\begin{align}
    \boldsymbol{j}^{\rm dia3} = \frac{e^3}{2V\hbar}\sum_{n, m,\omega',i, j,\boldsymbol{k}}(f_n(\boldsymbol{k},\mu)-f_m(\boldsymbol{k}, \mu))\nabla_{\boldsymbol{k}}\frac{1}{\varepsilon_{n}(\boldsymbol{k})-\varepsilon_{m}(\boldsymbol{k})+\hbar\omega'}v_{nm}^{i}(\boldsymbol{k})v_{mn}^{j}(\boldsymbol{k})A^{i}(\omega')A^{j}(-\omega').
    \label{j_dia3}
\end{align}
\end{widetext}
Similar to injection current, with time-reversal symmetry (TRS), this term is nonzero only with CP light illumination. Later we show how this term can be combined with $\boldsymbol{j}^{\rm off}$ and constitute the $\boldsymbol{j}^{\rm Fermi}$. 

We now switch to $\boldsymbol{j}^{\rm off}$. By replacing $\bm{v}_{mn}(\bm{k})$ with $\bm{r}_{mn}(\bm{k})$ according to (S6), $\boldsymbol{j}^{\rm off}$ is written as:
\begin{align}
        \boldsymbol{j}^{\rm off} = -\frac{e^3 i}{V\hbar}\sum_{l,m, n,i,j,\boldsymbol{k},\omega'} \frac{(f_{m}^{T}(\boldsymbol{k},\mu)-f_{l}^{T}(\boldsymbol{k},\mu))v^{j}_{nl}(\boldsymbol{k})v^{i}_{lm}(\boldsymbol{k})\boldsymbol{r}_{mn}(\boldsymbol{k})A_{i}(\omega')A_{j}(-\omega')}{\varepsilon_{l}(\boldsymbol{k})-\varepsilon_{m}(\boldsymbol{k})+\hbar\omega'-i\hbar\eta} \nonumber \\+ \frac{(f_{n}^{T}(\boldsymbol{k},\mu)-f_{l}^{T}(\boldsymbol{k},\mu))v^{i}_{nl}(\boldsymbol{k})v^{j}_{lm}(\boldsymbol{k})\boldsymbol{r}_{mn}(\boldsymbol{k})A_{i}(\omega')A_{j}(-\omega')}{\varepsilon_{n}(\boldsymbol{k})-\varepsilon_{l}(\boldsymbol{k})+\hbar\omega'-i\hbar\eta} \nonumber \\
        + \frac{e^3 i}{V\hbar}\lim_{\epsilon \xrightarrow{} 0 }\sum_{l,m=n,i,j,\boldsymbol{k},\omega'} \frac{(f_{m}^{T}(\boldsymbol{k},\mu)-f_{l}^{T}(\boldsymbol{k},\mu))v^{i}_{lm}(\boldsymbol{k})}{\varepsilon_{l}(\boldsymbol{k})-\varepsilon_{m}(\boldsymbol{k})+\hbar\omega'-i\hbar\eta} \int_{\boldsymbol{k}-\epsilon}^{\boldsymbol{k}+\epsilon}d\boldsymbol{k'}v^{\boldsymbol{k'}\boldsymbol{k},j}_{nl}\boldsymbol{r}_{mn}^{\boldsymbol{k}\boldsymbol{k'}}A_{i}(\omega')A_{j}(-\omega') \nonumber \\
        +\frac{(f_{n}^{T}(\boldsymbol{k},\mu)-f_{l}^{T}(\boldsymbol{k},\mu))v^{i}_{nl}(\boldsymbol{k})}{\varepsilon_{n}(\boldsymbol{k})-\varepsilon_{l}(\boldsymbol{k})+\hbar\omega'-i\hbar\eta}\int_{\boldsymbol{k}-\epsilon}^{\boldsymbol{k}+\epsilon}d\boldsymbol{k'}v^{\boldsymbol{k}\boldsymbol{k'},j}_{lm}\boldsymbol{r}_{mn}^{\boldsymbol{k'}\boldsymbol{k}}A_{i}(\omega')A_{j}(-\omega').
\end{align}
The first two lines are demonstrated to be zero by rotating $l\xrightarrow{} n$, $m\xrightarrow{}l$ and using commutation relation $[r^{i},P^{j}] = i\hbar\delta_{ij}$. Using (S7) and (S8), we get:

\begin{align}
    \boldsymbol{j}^{\rm off} = \frac{e^3 i}{V\hbar}\sum_{n,m,i,j,\omega',\boldsymbol{k}} \frac{(f_{m}^{T}(\boldsymbol{k},\mu)-f_{n}^{T}(\boldsymbol{k},\mu))}{\varepsilon_{n}(\boldsymbol{k})-\varepsilon_{m}(\boldsymbol{k})+\hbar\omega'-i\eta}A^{i}(\omega')A^{j}(-\omega')\nonumber
    \\\Big(v_{nm}^{i}(\boldsymbol{k})i\boldsymbol{\nabla_{\boldsymbol{k}}}{v_{mn}^{j}(\boldsymbol{k})} +(\boldsymbol{R}_{mm}(\boldsymbol{k})-\boldsymbol{R}_{nn}(\boldsymbol{k}))v_{nm}^{i}(\boldsymbol{k})v_{mn}^{j}(\boldsymbol{k})\Big).
\end{align}

Similar to diagonal parts, $j^{\rm off}$ can be separated as a resonant contribution related with $\delta$ function:
\begin{widetext}
\begin{align}
        \boldsymbol{j}^{\rm off1} =& -\frac{e^3\pi }{V\hbar}\sum_{n,m,i,j,\omega',\boldsymbol{k}} (f_{m}^{T}(\boldsymbol{k},\mu)-f_{n}^{T}(\boldsymbol{k},\mu))\Big(v_{nm}^{i}(\boldsymbol{k})i\boldsymbol{\nabla_{\boldsymbol{k}}}{v_{mn}^{j}(\boldsymbol{k})} +(\boldsymbol{R}_{mm}(\boldsymbol{k})-\boldsymbol{R}_{nn}(\boldsymbol{k}))v_{nm}^{i}(\boldsymbol{k})v_{mn}^{j}(\boldsymbol{k})\Big)\nonumber \\
        &\times \delta(\varepsilon_{n}(\boldsymbol{k})-\varepsilon_{m}(\boldsymbol{k})+\hbar\omega')A^{i}(\omega')A^{j}(-\omega'),
\label{j_off1}         
\end{align}
\end{widetext}
 and the remaining nonresonant contribution is related with principal part:
\begin{widetext}
\begin{align}
        \boldsymbol{j}^{\rm off2} =& \frac{e^3 i }{V\hbar}\sum_{n,m,i,j,\omega',\boldsymbol{k}} (f_{m}^{T}(\boldsymbol{k},\mu)-f_{n}^{T}(\boldsymbol{k},\mu))\Big(v_{nm}^{i}(\boldsymbol{k})i\boldsymbol{\nabla_{\boldsymbol{k}}}{v_{mn}^{j}(\boldsymbol{k})} +(\boldsymbol{R}_{mm}(\boldsymbol{k})-\boldsymbol{R}_{nn}(\boldsymbol{k}))v_{nm}^{i}(\boldsymbol{k})v_{mn}^{j}(\boldsymbol{k})\Big)\nonumber \\
        &\times\mathcal{P}\frac{1}{\varepsilon_{n}(\boldsymbol{k})-\varepsilon_{m}(\boldsymbol{k})+\hbar\omega'}A^{i}(\omega')A^{j}(-\omega').
\label{j_off2}        
\end{align}
\end{widetext}
With TRS, by switching $n\leftrightarrow m$ and $\boldsymbol{k} \leftrightarrow \boldsymbol{-k}$, $j^{\rm off2}$ vanishes with LP light illumination, and resonant $j^{\rm off1}$ remains. This is generally referred to as shift current ($\bm{j}^{\rm shift}$). With CP light, resonant $j^{\rm off1}$ vanishes, but nonresonant $j^{\rm off2}$ is nonvanishing. By switching $n\leftrightarrow m$, $i\leftrightarrow j$, and $\omega' \leftrightarrow -\omega'$, the term $\big(\boldsymbol{R}_{mm}(\mathbf{k})-\boldsymbol{R}_{nn}(\mathbf{k})\big)v_{nm}^{i}(\boldsymbol{k})v_{mn}^{j}(\boldsymbol{k})$  vanishes trivially. $j^{\rm off2}$ is reduced as:
\begin{widetext}
\begin{align}
    \boldsymbol{j}^{\rm off2} = \frac{e^3 }{2V\hbar}\sum_{n,m,\omega',i,j,\boldsymbol{k}}(f_{n}^{T}(\boldsymbol{k},\mu)-f_{m}^{T}(\boldsymbol{k},\mu))\frac{1}{\varepsilon_{n}(\boldsymbol{k})-\varepsilon_{m}(\boldsymbol{k})+\hbar\omega'}\nabla_{\boldsymbol{k}}(v_{nm}^{i}(\boldsymbol{k})v_{mn}^{j}(\boldsymbol{k}))A^{i}(\omega')A^{j}(-\omega').
\end{align}
\end{widetext}
Above we have inserted a factor of $\frac{1}{2}$ due to symmetrization by switching  $n\leftrightarrow{}m$, $i\leftrightarrow{}j$, and $\omega'\leftrightarrow{}-\omega'$. Inspecting (S23) and (S28), $\nabla \boldsymbol{k}$ are applied on different parts of the same equation. Adding them together is equivalent to applying $\nabla \boldsymbol{k}$ derivative on $(f_{n}^{T}(\boldsymbol{k},\mu)-f_{m}^{T}(\boldsymbol{k},\mu))$: 
\begin{widetext}
\begin{align}
    \boldsymbol{j}^{dia3}+\boldsymbol{j}^{\rm off2} = - \frac{e^3 }{2V\hbar}\sum_{n,m,\omega',i,j,\boldsymbol{k}}\nabla_{\boldsymbol{k}}(f_{n}^{T}(\boldsymbol{k},\mu)-f_{m}^{T}(\boldsymbol{k},\mu))\frac{1}{\varepsilon_{n}(\boldsymbol{k})-\varepsilon_{m}(\boldsymbol{k})+\hbar\omega'}v_{nm}^{i}(\boldsymbol{k})v_{mn}^{j}(\boldsymbol{k})A^{i}(\omega')A^{j}(-\omega').
\end{align}
\end{widetext}
In semiconductors, $f_{n}^{T}(\boldsymbol{k},\mu)-f_{m}^{T}(\boldsymbol{k},\mu)$ is a constant, so $ \boldsymbol{j}^{\rm dia3}+\boldsymbol{j}^{\rm off2}$ is zero. For metals, when $T \xrightarrow{}0$, $\nabla_{\boldsymbol{k}}(f_{n}^{T}(\boldsymbol{k},\mu)-f_{m}^{T}(\boldsymbol{k},\mu)) \xrightarrow{} \hbar \boldsymbol{v}_{nn}(\boldsymbol{k})\delta(\varepsilon_{n}(\boldsymbol{k})-\mu)-\hbar \boldsymbol{v}_{mm}(\boldsymbol{k})\delta(\varepsilon_{m}(\boldsymbol{k})-\mu)$. The current is contributed by excitations from both occupied states to Fermi surface and from Fermi surface to unoccupied states ($\bm{j}^{\rm Fermi}$). By switching $n \leftrightarrow m$, we rewrite (S29) as:
 \begin{widetext}
 \begin{align}
\bm{j}^{\rm Fermi}= -\dfrac{e^3}{V\hbar}\sum_{\bm{k},n,m,i,j,\omega'=\pm \omega}\bm{\nabla}_{\bm{k}}f_n^T(\bm{k},\mu) \dfrac{v_{nm}^i(\bm{k})v_{mn}^j(\bm{k})}{\varepsilon_n(\bm{k})-\varepsilon_m(\bm{k})+\hbar\omega'}A^i(\omega' )A^j(-\omega')
\label{j_Fermi_quadra}
 \end{align}
 \end{widetext}

\section{Gauge invariance of nonlinear DC photocurrent}
We add arbitrary phases to the periodic function part of Bloch states $u_{n}(\bm{k}) \xrightarrow{} e^{i\phi_n (\bm{k})}u_{n}(\bm{k})$, $u_{m}(\bm{k}) \xrightarrow{} e^{i\phi_m (\bm{k})}u_{m}(\bm{k})$ to see whether expressions of photocurrent will change.
Since velocity operator $\hat{\bm{v}}^{\bm{k}}$ does not enclose any phase information, velocity matrix element $\bm{v}_{nm}(\bm{k})$ will transform following:
\
\begin{equation}
        \bm{v}_{nm}(\bm{k})\xrightarrow{} e^{i(\phi_{m}(\bm{k}) - \phi_{n}(\bm{k}))}\bm{v}_{nm}(\bm{k}),
\end{equation}
    \begin{align} \bm{v}_{nm}^{i}(\bm{k})\bm{v}_{mn}^{j}(\bm{k})\xrightarrow{}e^{i\phi_{mn}(\bm{k})}\bm{v}_{nm}^{i}(\bm{k})e^{i\phi_{nm}(\bm{k})}\bm{v}_{mn}^{j}(\bm{k}) = \bm{v}_{nm}^{i}(\bm{k})\bm{v}_{mn}^{j}(\bm{k}). 
\end{align}
According to Eqs. \ref{j_dia1} and \ref{j_Fermi_quadra}, $\bm{j}^{\rm inj}$ and $\bm{j}^{\rm Fermi}$ are thus gauge invariant. 

Below we prove the gauge invariance of $\bm{j}^{\rm shift}$. By adding arbitrary phases, we have the following transformation:
\begin{align}
    \bm{v}_{nm}^{i}(\bm{k})\nabla_{\bm{k}}\bm{v}_{mn}^{j}(\bm{k}) \xrightarrow{} e^{i\phi_{mn}(\bm{k})}\bm{v}_{nm}^{i}(\bm{k})e^{i\phi_{nm}(\bm{k})}\nabla_{\bm{k}}\bm{v}_{mn}^{j}(\bm{k}) + e^{i\phi_{mn}(\bm{k})}\bm{v}_{nm}^{i}(\bm{k})e^{i\phi_{nm}(\bm{k})}\bm{v}_{mn}^{j}(\bm{k})\nabla_{\bm{k}}(i\phi_{nm}(\bm{k}))
    \nonumber\\
    =\bm{v}_{nm}^{i}(\bm{k})\nabla_{\bm{k}}\bm{v}_{mn}^{j}(\bm{k}) + \bm{v}_{nm}^{i}(\bm{k})\bm{v}_{mn}^{j}(\bm{k})\nabla_{\bm{k}}(i\phi_{nm}(\bm{k})),
\end{align}
\begin{align}
    \bm{R}_{mm}(\bm{k})-\bm{R}_{nn}(\bm{k}) \xrightarrow{} \bm{R}_{mm}(\bm{k}) + i 
    \braket{u_m(\bm{k})|u_m(\bm{k})}
    \nabla_{\bm{k}}(i\phi_{m}(\bm{k}))  - \bm{R}_{nn}(\bm{k}) - i 
    \braket{u_n(\bm{k})|u_n(\bm{k})}\nabla_{\bm{k}}(i\phi_{n}(\bm{k}))\nonumber \\
    = \bm{R}_{mm}(\bm{k}) - \bm{R}_{nn}(\bm{k}) + i\nabla_{\bm{k}}(i\phi_{mn}(\bm{k})),
\end{align}
\begin{align}
    i\bm{v}_{nm}^{i}(\bm{k})\nabla_{\bm{k}}\bm{v}_{mn}^{j}(\bm{k}) + (\bm{R}_{mm}(\bm{k})-\bm{R}_{nn}(\bm{k}))\bm{v}_{nm}^{i}(\bm{k})\bm{v}_{mn}^{j}(\bm{k})\xrightarrow{}
    i\bm{v}_{nm}^{i}(\bm{k})\nabla_{\bm{k}}\bm{v}_{mn}^{j}(\bm{k}) + (\bm{R}_{mm}(\bm{k})-\bm{R}_{nn}(\bm{k}))\nonumber \\
    \bm{v}_{nm}^{i}(\bm{k})\bm{v}_{mn}^{j}(\bm{k}).
\end{align}
Therefore, $\bm{j}^{\rm shift}$ is also gauge invariant.

\section{Nonlinear DC photocurrent:     Time-reversal symmetry broken}
 When TRS is broken in magnetic systems, the relation $\bm{v}_{nm}(\bm{k}) = -\bm{v}^{*}_{nm}(-\bm{k})$ will no longer hold. Here we reinspect the expressions for each contribution. 
 
 We look at the diagonal contribution $\bm{j}^{\rm dia}  =\frac{e}{V}\sum_{n, \bm{k}}{\bm{v}_{nn}(\bm{k})\rho_{nn}(\bm{k}})$ first. As shown in Eqs. \ref{dia_factor}-\ref{j_dia3}, it can be parsed into the resonant injection current $\bm{j}^{\rm dia1}$ ($\bm{j}^{\rm inj}$) and nonresonant $\bm{j}^{\rm dia2}$ and $\bm{j}^{\rm dia3}$.  For the resonant part, without $\bm{v}_{nn}(\bm{k}) \neq -\bm{v}_{nn}(-\bm{k})$, $\bm{j}^{\rm inj}$ can also survive with the illumination of LP light. This is referred to as ``magnetic injection current'' and has been computed recently~\cite{zhang2019switchable,fei2020giant}. $\bm{j}^{\rm dia2}$ vanishes trivially and the cancellation does not rely on TRS. The remaining $\bm{j}^{\rm dia3}$ is associated with the nonresonant part of $\bm{j}^{\rm off}  =\frac{e}{V}\sum_{n \neq  m, \bm{k}}{\bm{v}_{nm}(\bm{k})\rho_{mn}(\bm{k}})$ below and constitutes the $\bm{j}^{\rm Fermi}$.
 
 For the off-diagonal contribution $\bm{j}^{\rm off}  =\frac{e}{V}\sum_{n\neq m, \bm{k}}{\bm{v}_{mn}(\bm{k})\rho_{nm}(\bm{k}})$, as shown by Eq. \ref{j_off1}, $\bm{j}^{\rm shift}$ represents the resonant part. When  $\bm{v}_{nm}(\bm{k}) = -\bm{v}^{*}_{nm}(-\bm{k})$ does not hold, $\bm{j}^{\rm shift}$ is nonvanishing with both LP and CP light. This is in contrast with nonmagnetic systems where $\bm{j}^{\rm shift}$ is nonvanishing only with LP light. For the nonresonant part of $\bm{j}^{\rm off}$ (Eq. \ref{j_off2}), the term $\big(\bm{R}_{mm}(\bm{k})-\bm{R}_{nn}(\bm{k})\big)v_{nm}^{i}(\bm{k})v_{mn}^{j}(\bm{k})$ vanishes trivially independent of TRS.
 With TRS breaking, the term $v_{nm}^{i}(\bm{k})i\bm{\nabla}_{\bm{k}} v_{mn}^{j}(\bm{k})$ is not cancelled between $\bm{k}$ and $-\bm{k}$. Combined with $\bm{j}^{\rm dia3}$, it arrives in the form of Eq. \ref{j_Fermi_quadra} ($\bm{j}^{\rm Fermi}$) and is nonvanishing with both LP and CP light. 
 
 As a summary, by breaking TRS, the nonlinear DC photocurrent will have $\bm{j}^{\rm inj}$ (Eq. \ref{j_dia1}), $\bm{j}^{\rm shift}$ (Eq. \ref{j_off1}), and $\bm{j}^{\rm Fermi}$ (Eq. \ref{j_Fermi_quadra}) with illumination of both CP and LP light.  

\section{Non-Linear Photocurrents for General Bloch Hamiltonians}

Here we give an equivalent derivation of the above contributions to the non-linear DC photocurrent for a general Bloch Hamiltonian.  We show that in addition to the shift, injection, and $\bm{j}^{\bm Fermi}$ currents there is another contribution to the non-linear photocurrent that is only non-vanishing for systems with a Fermi surface.  We denote this contribution $\bm{J}_{\rm{Fermi}2}$.

\section{Time Evolution of the Quantum Density Matrix}

Here we solve for the nonlinear response of charge currents to quadratic order in a perturbing electric field by first solving the quantum density matrix, $\hat{\rho}(t)$, to quadratic order in the external electromagnetic vector potential.  To do so we employ the von Neumann equation which describes the time evolution of this quantum operator \cite{von1927wahrscheinlichkeitstheoretischer}.

\begin{equation}
i\hbar \dfrac{\partial \hat{\rho}(t)}{\partial t}=[\hat{H}(\hat{\bm{r}},\hat{\bm{p}},t),\hat{\rho}(t)]
\label{vonN}
\end{equation}

\noindent
The Hamiltonian, $\hat{H}(\hat{\bm{r}},\hat{\bm{p}},t)$, can be divided into two parts: $\hat{H}_0(\hat{\bm{r}},\hat{\bm{p}})$ and $\hat{H}'(\hat{\bm{r}},\hat{\bm{p}},t)$.  The time independent part, $\hat{H}_0(\hat{\bm{r}},\hat{\bm{p}})$, describes noninteracting electrons in a crystal potential, and the time dependent part describes the interaction between electrons in the material and the external perturbing electric field.  Due to the translation symmetry of the crystal it is advantageous to write the operators in the basis of  the  Bloch eigenstates of $\hat{H}_0(\hat{\bm{r}},\hat{\bm{p}})$:  $\mathcal{O}_{nm}(\bm{k},\bm{k}')=\bra{u_n(\bm{k})}\hat{\mathcal{O}}\ket{u_m(\bm{k}')}$, where $\ket{u_n(\bm{k})}$ is the periodic part of the Bloch wave function in band $n$, with crystal momentum $\bm{k}$, and energy $\varepsilon_n(\bm{k})$. Here we will study spatially homogeneous electric fields, $\bm{E}(\bm{r},t)\rightarrow \bm{E}(t)$, that only induce electronic transitions between Bloch states with the same crystal momentum $\bm{k}$ and thus, define $\mathcal{O}_{nm}(\bm{k},\bm{k}')=\mathcal{O}_{nm}(\bm{k},\bm{k}')\delta_{\bm{k},\bm{k}'}\rightarrow \mathcal{O}_{nm}(\bm{k})$.

The Bloch Hamiltonian $\hat{H}_0(\hat{\bm{r}},\hat{\bm{p}})$ can be written in the basis of localized atomic orbitals $\{\ket{\varphi_{\bm{R}_j}}\}$ whose matrix elements we denote as

\begin{equation}
H^0_{ij}(\bm{R}_i,\bm{R}_j')=\bra{\varphi_{\bm{R}_i}}\hat{H}_0(\hat{\bm{r}},\hat{\bm{p}})\ket{\varphi_{\bm{R}'_j}}
\end{equation}

\noindent
Here $\bm{R}_i$ indexes the position of orbital $i$ in the unit cell at $\bm{R}$.  Due to translation symmetry $H^0_{ij}(\bm{R}_i,\bm{R}_j')\rightarrow H^0_{ij}(\bm{R}_i-\bm{R}_j')$.  These orbitals are chosen such that they satisfy

\begin{equation}
\bm{R}_i=\bm{R}+\bm{\tau}_i=\bra{\varphi_{\bm{R}_i}}\hat{\bm{r}}\ket{\varphi_{\bm{R}_i}}
\end{equation}

\noindent
where $\bm{\tau}_i$ is the intracellular position of orbital $i$.  We can then write the operator as 

\begin{equation}
\hat{H}_0(\hat{\bm{r}},\hat{\bm{p}})=\sum \limits_{\bm{R},\bm{R}',i,j}\hat{c}^\dagger_{\varphi}(\bm{R}_i)H^0_{ij}(\bm{R}_i,\bm{R}_j')\hat{c}_\varphi(\bm{R}'_j)
\end{equation}

\noindent
where $\hat{c}^\dagger_{\varphi}(\bm{R}_i)$ and $\hat{c}_\varphi(\bm{R}_i)$ are electronic creation and annihilation operators for the $i$ orbital in the unit cell at $\bm{R}$.

To make connection with our Bloch states, we can build Bloch like function out of these localized atomic orbitals by taking the linear combination

\begin{align}
\braket{\bm{r}|\chi^{\bm{k}}_i}=\dfrac{1}{\sqrt{N}}\bra{\bm{r}}\sum_{\bm{R}} e^{i\bm{k}\cdot\hat{\bm{r}}}\ket{\varphi_{R_i}} \nonumber\\
=\dfrac{1}{\sqrt{N}}\sum_{\bm{R}} e^{i\bm{k}\cdot (\bm{R}+\bm{\tau}_i)}\braket{\bm{r}|\varphi_{R_i}}\label{phaseconvention}
\end{align}

\noindent
We can define our Bloch Hamiltonian and its components by taking the inner product of the Hamiltonian with respect to these states

\begin{widetext}
\begin{align}
H^0_{ij}(\bm{k})&=\bra{\chi^{\bm{k}}_i}\hat{H}_0(\hat{\bm{r}},\hat{\bm{p}})\ket{\chi^{\bm{k}}_j}=\sum_{\bm{R},\bm{R}'}\dfrac{1}{N}e^{-i\bm{k}\cdot(\bm{R}_i-\bm{R}'_j+\bm{\tau}_i-\bm{\tau}_j)}\bra{\varphi_{\bm{R}_i}}\hat{H}_0(\hat{\bm{r}},\hat{\bm{p}})\ket{\varphi_{\bm{R}_j}}  \nonumber\\
& = \sum_{\tilde{\bm{R}}}e^{-i\bm{k}\cdot(\tilde{\bm{R}}+\bm{\tau}_i-\bm{\tau}_j)}\bra{\varphi_{\bm{0}_i}}\hat{H}_0(\hat{\bm{r}},\hat{\bm{p}})\ket{\varphi_{\tilde{\bm{R}}_j}}=\sum_{\tilde{\bm{R}}}e^{-i\bm{k}\cdot(\tilde{\bm{R}}+\bm{\tau}_i-\bm{\tau}_j)}H^0_{ij}(\tilde{\bm{R}})
\label{blochH}
\end{align}
\end{widetext}

\noindent
Here $\tilde{\bm{R}}=\bm{R}_i-\bm{R}_j$ and note that $H^0_{ij}(\bm{R})$ is periodic under translation by a lattice vector.  We can diagonalize $H^0_{ij}(\bm{k})$ such that 

\begin{equation}
\sum_j H^0_{ij}(\bm{k})u_{n\bm{k}}(j)=\varepsilon_n(\bm{k})u_{n\bm{k}}(i)
\end{equation}

\noindent
These coefficients $u_{n\bm{k}}(i)$ make up the components of the periodic part of the Bloch vectors $\ket{u_n(\bm{k})}$ written in the orbital basis.

Light matter interaction between electrons in the material and an external electromagnetic field can be taken into account via the Peierls substitution \cite{peierls1933theorie}

\begin{equation}
 \hat{c}^\dagger_{\varphi}(\bm{R}_i) \hat{c}_{\varphi}(\bm{R}'_j)\rightarrow  \hat{c}^\dagger_{\varphi}(\bm{R}_i) \hat{c}_{\varphi}(\bm{R}'_j)e^{-ie/\hbar\int^{\bm{R}+\bm{\tau}_i}_{\bm{R}'+\bm{\tau}_j}\bm{A}_{ext}(\bm{r}',t)\cdot d\bm{r}'}
\end{equation}

\noindent
Here we will work in temporal gauge such that the external electromagnetic scalar potential $\phi_{ext}(\bm{r},t)$ can be taken to vanish.  For spatially homogeneous electromagnetic vector fields, $\bm{A}(t)$, we can expand the Hamiltonian in the orbital basis as

\begin{align}
\hat{H}(t)=\sum \limits_{\bm{R},\bm{R}',i,j}\hat{c}^\dagger_{\varphi}(\bm{R}_i)H^0_{ij}(\bm{R}_i,\bm{R}_j')\hat{c}_\varphi(\bm{R}'_j)
 e^{-ie/\hbar ((\bm{R}+\bm{\tau}_i)-(\bm{R}'+\bm{\tau}_j))\cdot\bm{A}_{ext}(t)}
\end{align}

\noindent
The external field has modified the matrix elements of the Hamiltonian such that 

\begin{equation}
H^0_{ij}(\bm{R}_i,\bm{R}_j')\rightarrow H^0_{ij}(\bm{R}_i,\bm{R}_j')e^{-ie/\hbar ((\bm{R}+\bm{\tau}_i)-(\bm{R}'+\bm{\tau}_j))\cdot\bm{A}_{ext}(t)}
\end{equation}

\noindent
Using equation \ref{blochH} it can be shown that in the Bloch basis 

\begin{equation}
H^0_{ij}(\bm{k})\rightarrow H_{ij}(\bm{k},t)=H^0_{ij}(\bm{k}+\dfrac{e}{\hbar}\bm{A}(t))
\label{modH}
\end{equation}

\noindent
We see that the introduction of spatially homogeneous electromagnetic vector potential has simply boosted the momentum of the matrix elements of the Bloch Hamiltonian.

To extract $\hat{H}'(\hat{\bm{r}},\hat{\bm{p}},t)$ one only needs to expand $\hat{H}(\hat{\bm{r}},\hat{\bm{p}},t)$ in a power series of the electromagnetic vector potential $\bm{A}(t)$.  By substitution into the von Neumann equation we can begin to solve for the density matrix in powers of the external electromagnetic field with the identification $\bm{E}(t)=-\partial_t \bm{A}(t)$ made possible by use of the temporal gauge. 

The von Neumann equation is a first order linear differential equation in time.  To solve the equation an initial condition on the density matrix must be specified.  Before application of the external electromagnetic field the material is in its ground state and can be describe by $f^T_n(\bm{k},\mu)$ the Fermi occupation function describing the probability for an electron to be in the Bloch state in band $n$ at momentum $\bm{k}$ given the temperature $T$ and chemical potential $\mu$ of the system.

Here we will mainly be concerned with materials perturbed by light with a few nonzero Fourier components $\omega$.  We thus Fourier transform equation \ref{vonN} and write it in the Bloch basis

\begin{widetext}
\begin{equation}
-\hbar\omega \rho_{nm}(\bm{k},\omega)=\sum_{l,\omega'}\bigg(H_{nl}(\bm{k},\omega')\rho_{lm}(\bm{k},\omega-\omega')-\rho_{nl}(\bm{k},\omega')H_{lm}(\bm{k},\omega-\omega')\bigg)
\label{vonE}
\end{equation}
\end{widetext}

\noindent
where $\rho_{nm}(\bm{k},\omega)=\int dt \rho_{nm}(\bm{k},t)e^{-i\omega t}$ and

\begin{equation}
H_{nm}(\bm{k},\omega)=\int dt e^{-i\omega t}\sum \limits_{ij}u^*_{\bm{k}n}(i)H^0_{ij}(\bm{k}+e/\hbar \bm{A}(t))u_{\bm{k}m}(j)
\end{equation}

\noindent
is the Fourier transform of the perturbed Bloch Hamiltonian written in the basis of eigenstates of $H^0_{ij}(\bm{k})$.  In doing so our differential equation has become an algebraic equation one can readily solve in powers of $\bm{E}(t)$.

\section{Quantum Current Operator}

Here we wish to determine the induced charge current in a material order by order in the perturbing external electric field.  This can be calculated with knowledge of $\hat{\rho}(t)$ and the quantum current operator $\hat{\bm{j}}(t)$ via

\begin{align}
\bm{j}(t)=\dfrac{1}{V}\text{Tr}[e\hat{\bm{v}}(t)\hat{\rho}(t)]
\end{align}

\noindent
We consider perturbation from spatially homogeneous electric fields that gives rise to spatially homogeneous currents.  We thus can identify the current operator with the product of the electric charge $e$ and the velocity operator $\hat{\bm{v}}(t)$ whose general form is $\hat{\bm{v}}(t)=i[\hat{H}(\hat{\bm{r}},\hat{\bm{p}},t),\hat{\bm{r}}]/\hbar$.

Due to the translation symmetry of the crystal it is convenient to express the velocity operator in the basis of our Bloch like functions $\ket{\chi^{\bm{k}}_i}$.  We take the inner product of the operator with respect to two such states and find

\begin{widetext}
\begin{align}
\bm{v}_{ij}(\bm{k},t)&=\bra{\chi^{\bm{k}}_i}\hat{\bm{v}}(t)\ket{\chi^{\bm{k}}_j}=\dfrac{i}{\hbar}\bra{\chi^{\bm{k}}_i}[\hat{H}(\hat{\bm{r}},\hat{\bm{p}},t),\hat{\bm{r}}]\ket{\chi^{\bm{k}}_j}  \nonumber \\
&=\dfrac{i}{\hbar N}\sum_{\bm{R},\bm{R}'}\bra{\varphi_{R_i}}\hat{H}(\hat{\bm{r}},\hat{\bm{p}},t)\ket{\varphi_{R'_j}}e^{-i\bm{k}\cdot(\bm{R}-\bm{R}'+\bm{\tau}_i-\bm{\tau}_j)} (\bm{R}'-\bm{R}+\tau_j-\tau_i) \nonumber \\
&=\dfrac{1}{\hbar}\bm{\nabla}_{\bm{k}} H_{ij}(\bm{k},t)
\label{vop}
\end{align}
\end{widetext}

\noindent
It is important to recognize that the representation of the velocity operator on Bloch states only takes this form in the gauge choice on $\ket{\chi_j^{\bm{k}}}$ shown above.  With this gauge choice we choose the Bloch state

\begin{align}
\ket{\psi_{nk}}
=\sum_R\sum_j u_{n\bm{k}}(j) e^{i\bm{k}\cdot(\bm{R}+\bm{\tau}_j)}\ket{\varphi_{R_j}}
\end{align}

\noindent
to have the property $\ket{\psi_{n (\bm{k}+\bm{G})}}=\ket{\psi_{n\bm{k}}}$ which implies that $u_{n(\bm{k}+\bm{G})}(j)=e^{-i\bm{G}\cdot\bm{\tau}_j}u_{n\bm{k}}(j)$.  This is indeed consistent with how $H_{ij}(\bm{k})$ is defined as $H_{ij}(\bm{k}+\bm{G})=e^{-i\bm{k}\cdot\bm{\tau}_i}H_{ij}(\bm{k})e^{i\bm{k}\cdot\bm{\tau}_j}$.

Here we will be considering spatially homogeneous electric fields with few nonzero frequency components $\omega$, and thus focus on the Fourier transform of the current

\begin{equation}
\bm{j}(\omega)=\dfrac{1}{V}\sum_{\omega'}\text{Tr}[e\hat{\bm{v}}(\omega-\omega')\hat{\rho}(\omega')]
\label{cur}
\end{equation}

\noindent
Again due to the translation symmetry of the crystal, it is convenient to compute the trace of these operators in the Bloch basis.  With knowledge of the representation of $\hat{\bm{v}}(\omega)$ and $\hat{\rho}(\omega)$ in the Bloch basis, equation \ref{cur} can be solve order by order in the electric field to compute the induced current.

\section{Charge Currents First Order in the Electric Field}

To solve for charge currents linear in the electric field we first expand the quantum density matrix and velocity operator in a power series with respect to $\bm{A}_{ext}(t)$

\begin{align}
\rho_{nm}(\bm{k},\omega)=\sum_p \rho^{(p)}_{nm}(\bm{k},\omega) \nonumber \\
\bm{v}_{nm}(\bm{k},\omega)=\sum_p \bm{v}^{(p)}_{nm}(\bm{k},\omega)
\end{align}

\noindent
where $p$ indexes the order to which $\rho_{nm}^{(p)}(\bm{k},\omega)$ and $\bm{v}^{(p)}_{nm}(\bm{k},\omega)$
is proportional to $\bm{A}(t)$.  There will be two contributions to the induced current at this order in the electric field.  We write the total first order current $\bm{j}^{(1)}(\omega)$ as the sum of these two contributions $\bm{j}_A^{(1)}(\omega)$ and $\bm{j}_B^{(1)}(\omega)$.  The first derives from the first order contribution $\rho_{nm}^{(1)}(\bm{k},\omega)$ to the quantum density matrix traced against the zeroth order contribution to the velocity operator $\hat{\bm{v}}^{(0)}(\omega)$.   The second comes from the zeroth order contribution to the density operator $\rho_{nm}^{(0)}(\bm{k},\omega)$ traced against the first order contribution to the velocity operator $\hat{\bm{v}}^{(1)}(\omega)$.

\begin{align}
  \bm{j}_A^{(1)}(\omega)&=\sum_{\omega'}\text{Tr}[e\hat{\bm{v}}^{(1)}(\omega-\omega')\hat{\rho}^{(0)}(\omega')] \nonumber \\
 \bm{j}_B^{(1)}(\omega)&=\sum_{\omega'}\text{Tr}[e\hat{\bm{v}}^{(0)}(\omega-\omega')\hat{\rho}^{(1)}(\omega')]
  \label{cur1}
\end{align}

\sloppypar{
\noindent
Using equation \ref{vop} and equation \ref{modH} the zeroth and first order terms to the velocity operators written in the Bloch basis are $\bm{v}_{nm}^{(0)}(\bm{k},\omega)=\delta_{0,\omega}/\hbar\bra{u_n(\bm{k})}\bm{\nabla}_{\bm{k}}\hat{H}_0(\bm{k})\ket{u_m(\bm{k})}$ and $\bm{v}_{nm}^{(1)}(\bm{k},\omega)=\sum \limits_ie/\hbar^2\bra{u_n(\bm{k})}\bm{\nabla}_{\bm{k}}\partial_{k_i}\hat{H}_0(\bm{k})\ket{u_m(\bm{k})}A_i(\omega)$ respectively.  The zeroth order contribution to the density matrix $\rho_{nm}^{(0)}(\bm{k},\omega)$ describes the occupation of the electrons in the ground state before the external electric field is applied to the material, which in the Bloch basis is simply the Fermi occupation functions
}

\begin{equation}
 \rho_{nm}^{(0)}(\bm{k},\omega)=\delta_{nm}\delta_{\omega,0}f_n^T(\bm{k},\mu)
 \label{rho0}
\end{equation}

\noindent
We see that the first contribution to equation \ref{cur1} can be simply written as

\begin{equation}
 \bm{j}_A^{(1)}(\omega)=\sum \limits_{i,n}\dfrac{e^2A_i(\omega)}{V\hbar^2}f_n^T(\bm{k},\mu)\bra{u_n(\bm{k})}\bm{\nabla}_{\bm{k}}\partial_{k_i}\hat{H}_0(\bm{k})\ket{u_n(\bm{k})}
 \end{equation}

To derive the contribution to the current proportional to $\bm{v}_{nm}^{(0)}(\bm{k},\omega)$ we must solve the von Neumann equation to obtain $\rho_{nm}^{(1)}(\bm{k},\omega)$.  Equating first order contributions in equation \ref{vonE} we have

\begin{widetext}
\begin{align}
-\hbar\omega \rho^{(1)}_{nm}(\bm{k},\omega)=(\varepsilon_n(\bm{k})-\varepsilon_m(\bm{k}))\rho_{nm}^{(1)}(\omega)+e\sum_{l,\omega'}\bigg(\bm{A}(\omega')\cdot\bm{v}_{nl}(\bm{k},\omega')\rho^{(0)}_{lm}(\bm{k},\omega-\omega')-\rho^{(0)}_{nl}(\bm{k},\omega-\omega')\bm{v}_{nl}(\bm{k},\omega')\cdot\bm{A}(\omega')\bigg)
\end{align}
\end{widetext}

\noindent
Where we have used equation \ref{modH} to expand the Bloch Hamiltonian in powers of $\bm{A}(\omega)$.  Using equation \ref{rho0}, we can substitute $\rho_{nm}^{(0)}(\bm{k},\omega)$ into the above and solve for $\rho_{nm}^{(1)}(\bm{k},\omega)$.  

\begin{align}
\rho_{nm}^{(1)}(\bm{k},\omega)=\dfrac{e(f_n^T(\bm{k},\mu)-f_m^T(\bm{k},\mu))\bm{v}_{nm}(\bm{k})\cdot \bm{A}(\omega)}{\varepsilon_n(\bm{k})-\varepsilon_m(\bm{k})+\hbar\omega}
\label{order1rho}
\end{align}

\noindent
Where we have defined the unperturbed velocity operator $\bm{v}_{nm}(\bm{k})=1/\hbar\bra{u_n(\bm{k})}\bm{\nabla}_{\bm{k}}\hat{H}_0(\bm{k})\ket{u_m(\bm{k})}$.  Tracing $\hat{\rho}^{(1)}(\bm{k},\omega)$ against $\hat{\bm{v}}^{(0)}(\bm{k},\omega)$ gives the second contribution to $\bm{j}^{(1)}(\omega)$.  

By expressing the electromagnetic vector potential $\bm{A}(\omega)$ in terms of the electric field $\bm{E}(\omega)$ we find the total contribution to $\bm{j}^{(1)}(\omega)$ can then be written as

\begin{align}
\bm{j}^{(1)}(\omega)=\dfrac{ie^2}{\omega V}\bigg[\sum \limits_{\bm{k},n,i}\bra{u_n(\bm{k})}\bm{\nabla}_{\bm{k}}\dfrac{\partial_{k_i}\hat{H}_0(\bm{k})}{\hbar^2}\ket{u_n(\bm{k})}f_n^T(\bm{k},\mu)\nonumber \\
+ \sum_{\bm{k},n,m,i} \dfrac{(f_n^T(\bm{k},\mu)-f_m^T(\bm{k},\mu))\bm{v}_{nm}(\bm{k})v^i_{mn}(\bm{k})}{\varepsilon_n(\bm{k})-\varepsilon_m(\bm{k})+\hbar\omega}\bigg]E_i(\omega) \hspace{.5cm}
\label{order1cur}
\end{align}

\noindent
Integration by parts in the first term of equation \ref{order1cur} will lead to a term proportional to the Fermi surface and other terms proportional to matrix elements of the dipole operator $\bm{R}_{nm}(\bm{k})=\bra{u_n(\bm{k})}i\bm{\nabla}\ket{u_m(\bm{k})}$.  Note that the representation of the $\hat{\bm{r}}$ on the period part of the Bloch functions is $i\bm{\nabla}$.  These extra terms, when combined with the second term in equation \ref{order1cur} resolve the divergence in the term, as $\omega \rightarrow 0$, stemming from $\bm{A}(\omega)=i \bm{E}(\omega)/\omega$.  This integration by parts and subsequent cancellation leads to an expression matching the canonical equation found when computing the first order current $\bm{j}^{(1)}(\omega)$ in {\it length} gauge \cite{aversa1995nonlinear,passos2018nonlinear}.

\section{Charge Currents Second Order in the Electric Field}

At second order in the electric field, the current $\bm{j}^{(2)}(\omega)$ has three unique contributions deriving from different orders in the power series expansions of the velocity operator and quantum density matrix.  We can write them as

\begin{align}
  \bm{j}_A^{(2)}(\omega)&=\sum_{\omega'}\text{Tr}[e\hat{\bm{v}}^{(2)}(\omega-\omega')\hat{\rho}^{(0)}(\omega')]  \\
 \bm{j}_B^{(2)}(\omega)&= \sum_{\omega'}\text{Tr}[e\hat{\bm{v}}^{(1)}(\omega-\omega')\hat{\rho}^{(1)}(\omega')] \\
   \bm{j}_C^{(2)}(\omega)&=\sum_{\omega'}\text{Tr}[e\hat{\bm{v}}^{(0)}(\omega-\omega')\hat{\rho}^{(2)}(\omega')]
  \label{cur2}
\end{align}
 
 \noindent
The second order contribution to the velocity operator written in the Bloch basis is 

\begin{align}
v^{(2)}_{nm}(\bm{k},\omega)&=\sum_{i,j,\omega'}\bra{u_n(\bm{k})}\bm{\nabla}\partial_{k_i}\partial_{k_j}\hat{H}_0(\bm{k})\ket{u_m(\bm{k})}
 \dfrac{e^2A_i(\omega-\omega')A_j(\omega')}{2\hbar^3}
\end{align}

\noindent
With knowledge of $v^{(1)}_{nm}(\bm{k})$ and $v^{(2)}_{nm}(\bm{k},\omega)$, and using equations \ref{order1rho} and \ref{rho0} the first two contributions $\bm{j}_A^{(2)}(\omega)$ and $\bm{j}_B^{(2)}(\omega)$ are

\begin{widetext}
\begin{align}
\bm{j}_A^{(2)}(\omega)&=\sum_{\bm{k},n,i,j,\omega'}\dfrac{e^3}{2V\hbar^3}f_n^T(\bm{k},\mu)\bra{u_n(\bm{k})}\bm{\nabla}_{\bm{k}}\partial_{k_i}\partial_{k_j}\hat{H}_0(\bm{k})\ket{u_n(\bm{k})} A_i(\omega')A_j(\omega-\omega')
\label{jA2}
\\
\bm{j}_B^{(2)}(\omega)&=\sum_{\bm{k},n,m,i,j,\omega'}\dfrac{e^3}{V\hbar^2}\dfrac{(f_n^T(\bm{k},\mu)-f_m^T(\bm{k},\mu))v^i_{nm}(\bm{k})}{\varepsilon_n(\bm{k})-\varepsilon_m(\bm{k})+\hbar\omega'}\bra{u_m(\bm{k})}\bm{\nabla}_{\bm{k}}\partial_{k_j}\hat{H}_0(\bm{k})\ket{u_n(\bm{k})}A_i(\omega')A_j(\omega-\omega')
\label{ord2curAB}
\end{align}
\end{widetext}

\noindent
To derive the last contribution to the second order current $\bm{j}_C^{(2)}(\omega)$ we first solve for $\hat{\rho}^{(2)}(\omega')$.  We do this in the Bloch basis by collecting terms that are second order in the electric field in equation \ref{vonE}.

\begin{widetext}
\small
\begin{gather}
-\hbar\omega \rho^{(2)}_{nm}(\bm{k},\omega)=(\varepsilon_n(\bm{k})-\varepsilon_m(\bm{k}))\rho_{nm}^{(2)}(\omega)+e\sum_{l,\omega'}\bigg(\bm{A}(\omega')\cdot\bm{v}_{nl}(\bm{k},\omega')\rho^{(1)}_{lm}(\bm{k},\omega-\omega')-\rho^{(1)}_{nl}(\bm{k},\omega-\omega')\bm{v}_{lm}(\bm{k},\omega')\cdot\bm{A}(\omega')\bigg) \nonumber \\
+\sum_{l,i,j,\omega',\omega''}\dfrac{e^2A_i(\omega'-\omega'')A_j(\omega'')}{2\hbar^2}\bigg(\bra{u_n(\bm{k})}\partial_{k_i}\partial_{k_j}\hat{H}_0(\bm{k})\ket{u_l(\bm{k})}
 \rho^{(0)}_{lm}(\bm{k},\omega-\omega')-\rho^{(0)}_{nl}(\bm{k},\omega-\omega')\bra{u_l(\bm{k})}\partial_{k_i}\partial_{k_j}\hat{H}_0(\bm{k})\ket{u_m(\bm{k})}\bigg)
 \label{vonN2}
\end{gather}
\normalsize
\end{widetext}

\noindent
Later we will find it useful to breakup the contributions to $\rho^{(2)}_{nm}(\bm{k},\omega)$ into two pieces. The first piece $\rho^{(2),0}_{nm}(\bm{k},\omega)$ is proportional to $\rho^{(0)}_{nm}(\bm{k},\omega)$ and whose contribution to the current we denote $\bm{j}_{C0}^{(2)}(\omega)$.  The second piece $\rho^{(2),1}_{nm}(\bm{k},\omega)$ is proportional to $\rho^{(1)}_{nm}(\bm{k},\omega)$ and whose contribution to the current we denote $\bm{j}_{C1}^{(2)}(\omega)$.  The total contribution to $\bm{j}_{C}^{(2)}(\omega)$ can thus be written as

\begin{align}
\bm{j}_{C}^{(2)}(\omega)&=\bm{j}_{C0}^{(2)}(\omega)+\bm{j}_{C1}^{(2)}(\omega)
=\sum_{\bm{k},n,m}(\bm{v}_{nm}(\bm{k})\rho^{(2),0}_{mn}(\bm{k},\omega)+\bm{v}_{nm}(\bm{k})\rho^{(2),1}_{mn}(\bm{k},\omega))
\end{align}

\section{DC Currents Second Order in the Electric Field}

Here we focus on the zero frequency contribution to the second order current $\bm{j}^{(2)}(0)$.  We will show how the contributions to $\bm{j}^{(2)}(0)$ can be combined in this DC limit such that they reproduce the established {\it shift} and {\it injection} current contributions to this nonlinear DC current.  In doing so we will show that the principle parts of the resonant pieces of the propagators in the expressions for $\bm{j}^{(2)}(0)$ vanish leaving contributions that are constrained to resonant transitions in the Brillouin zone and transitions on the Fermi surface.  

We start by analyzing $\bm{j}^{(2)}_{C1}(0)$.  We break this contribution into pieces proportional to the diagonal elements of $\rho^{(2),1}_{nm}(\bm{k},\omega)$ we denote $\bm{j}^{(2)}_{C1A}(0)$ and pieces proportional to the off diagonal elements of $\rho^{(2),1}_{nm}(\bm{k},\omega)$ we denote $\bm{j}^{(2)}_{C1B}(0)$.  To regulate the fermionic propagators we move the resonances off the real axes by taking $\hbar\omega'\rightarrow \hbar\omega'+i\eta$ for all frequencies.  At the end of the calculation we will take the limit as $\eta\rightarrow0$.  Using equation \ref{vonN2} we find

\begin{align}
\bm{j}^{(2)}_{C1A}(0)&=\sum_{\bm{k},i,j,n,m,\omega'=\pm \omega}\dfrac{e^3}{V} \dfrac{(f_n^T(\bm{k},\mu)-f_m^T(\bm{k},\mu))v^i_{nm}(\bm{k})v^j_{mn}(\bm{k})}{2i\hbar\eta(\varepsilon_n(\bm{k})-\varepsilon_m(\bm{k})+\hbar\omega'+i\hbar\eta)}
(\bm{v}_{nn}(\bm{k})-\bm{v}_{mm}(\bm{k}))A^i(\omega' )A^j(-\omega')
\label{curC1B}
\end{align}

\noindent
We now investigate the denominator on the right hand side of equation \ref{curC1B}.  In the limit as $\eta\rightarrow0$ we have

\begin{widetext}
\begin{align}
\lim_{\eta\rightarrow0}\dfrac{1}{2i\eta}\dfrac{1}{\varepsilon_n(\bm{k})-\varepsilon_m(\bm{k})+\hbar\omega'+i\hbar\eta}=\lim_{\eta\rightarrow0}\bigg(-\dfrac{\pi}{2\eta}\delta(\varepsilon_n(\bm{k})-\varepsilon_m(\bm{k})
+\hbar\omega') \nonumber \\
+\dfrac{1}{2i\eta}\mathcal{P}\dfrac{1}{\varepsilon_n(\bm{k})-\varepsilon_m(\bm{k})+\hbar\omega'}-\dfrac{1}{2}\mathcal{P}\dfrac{1}{(\varepsilon_n(\bm{k})-\varepsilon_m(\bm{k})+\hbar\omega')^2}\bigg)
\end{align}
\end{widetext}

\noindent
The second contribution to the denominator can be shown to vanish by relabeling the dummy indices in the sum of equation \ref{curC1B} ($i\leftrightarrow j$, $\omega' \leftrightarrow -\omega'$, and $n \leftrightarrow m$) leaving two contributions to $\bm{j}^{(2)}_{C1A}(0)$.

Now we investigate $\bm{j}^{(2)}_{C1B}(0)$, the contribution to the current $\bm{j}^{(2)}_{C1}(0)$ proportional to off diagonal elements of $\rho^{(2),1}_{nm}(\bm{k},\omega')$.  Again using equation \ref{vonN2} we find
\begin{gather}
    \rho_{nm}^{(2),1}(\bm{k},0) = -\sum_{i,j,l,\omega'=\pm \omega}\frac{e(\rho_{lm}^{(1)}(\bm{k},\omega')v_{nl}^{j}(\bm{k}) - \rho_{nl}^{(1)}(\bm{k},\omega')v_{lm}^{j}(\bm{k}))A^{j}(-\omega')}{\varepsilon_{n}(\bm{k}) - \varepsilon_{m}(
    \bm{k}) }\nonumber \\
    = - \frac{e^2}{\varepsilon_{n}(\bm{k})-\varepsilon_{m}(\bm{k})}\sum_{i,j,l,\omega'=\pm \omega}\Big(\frac{(f_l^T(\bm{k},\mu)-f_m^T(\bm{k},\mu)v_{nl}^{j}(\bm{k})v_{lm}^{i}(\bm{k})}{\varepsilon_{l}(\bm{k})-\varepsilon_{m}(\bm{k})+\hbar\omega'}
    - \frac{(f_n^T(\bm{k},\mu)-f_l^T(\bm{k},\mu))v_{nl}^{i}(\bm{k})v_{lm}^{j}(\bm{k})}{\varepsilon_{n}(\bm{k})-\varepsilon_{l}(\bm{k})+\hbar\omega'}\Big)A_{i}(\omega')A_{j}(-\omega').
\end{gather}
Switching $n\leftrightarrow{} l$ and $m\leftrightarrow{} l$, and using the relationship between off diagonal elements of the velocity operator and off diagonal elements of the dipole operator: $\bm{v}_{nm}(\bm{k})/(\varepsilon_n(\bm{k})-\varepsilon_m(\bm{k})+i\eta)=i\bm{R}_{nm}(\bm{k})/\hbar$ for $m\neq n$ and $\eta\rightarrow0$, we get
\begin{gather}
    \bm{j}^{(2)}_{C1B}(0)= \frac{e^{3}i}{V\hbar}\sum_{i,j,\omega'=\pm \omega}\Big(\sum_{l\neq m}\frac{(f_n^T(\bm{k},\mu)-f_m^T(\bm{k},\mu))v_{nm}^{i}(\bm{k})v_{ln}^{j}(\bm{k})\bm{R}_{ml}(\bm{k})}{\varepsilon_{l}(\bm{k})-\varepsilon_{m}(\bm{k})+\hbar\omega' +i\hbar\eta}
     \nonumber\\ - \sum_{l\neq n}\frac{(f_n^T(\bm{k},\mu)-f_m^T(\bm{k},\mu))v_{nm}^{i}(\bm{k})v_{ml}^{j}(\bm{k})\bm{R}_{ln}(\bm{k})}{\varepsilon_{l}(\bm{k})-\varepsilon_{m}(\bm{k})+\hbar\omega'+i\hbar\eta}\Big)A_i(\omega')A_j(-\omega').
\end{gather}
Using $\sum_{l}\ket{u_l(\bm{k})}\bra{u_l(\bm{k})} = \bm{1}$ and $\bra{u_n(\bm{k})}\partial_{k_i}\ket{u_m(\bm{k})}=-(\partial_{k_i}\bra{u_n(\bm{k})})\ket{u_m(\bm{k})}$ deriving from the orthogonality of the Bloch states, we get
\begin{equation}
    \sum_{l\neq m}\bm{R}_{ml}(\bm{k})v_{ln}^{j}(\bm{k}) = -i\bm{\nabla}_{\bm{k}}(\bra{u_m(\bm{k})})\hat{v}_j(\bm{k})\ket{u_n{\bm{k}}} - \bm{R}_{mm}(\bm{k})v_{mn}^{j}(\bm{k}) \nonumber,
\end{equation}
\begin{equation}
    \sum_{l\neq n}v_{ml}^{j}(\bm{k})\bm{R}_{ln}(\bm{k}) = i\bra{u_m(\bm{k})}\hat{v}_j(\bm{k})\bm{\nabla}_{\bm{k}}\ket{u_n{\bm{k}}} - \bm{R}_{nn}(\bm{k})v_{mn}^{j}(\bm{k}).
\end{equation}

We arrive at
\begin{align}
\bm{j}^{(2)}_{C1B}(0)=\sum_{\bm{k},n,m,i,j,\omega'=\pm \omega}\dfrac{e^3}{V\hbar}\dfrac{(f_n^T(\bm{k},\mu)-f_m^T(\bm{k},\mu))v_{nm}^i(\bm{k})}{\varepsilon_n(\bm{k})-\varepsilon_m(\bm{k})+\hbar\omega'+i\hbar\eta} \nonumber\\
\times \bigg(\bra{u_m(\bm{k})}\hat{v}_j(\bm{k})\bm{\nabla}_{\bm{k}}\ket{u_n{\bm{k}}}+(\bm{\nabla}_{\bm{k}}\bra{u_m(\bm{k})})\hat{v}_j(\bm{k})\ket{u_n({\bm{k}})} \nonumber \\
+i v_{mn}^j(\bm{k})(\bm{R}_{nn}(\bm{k})-\bm{R}_{mm}(\bm{k}))\bigg)A^i(\omega')A^j(-\omega').
\label{jc1B}
\end{align}

\noindent

We now add $\bm{j}^{(2)}_{C1B}(0)$ to $\bm{j}^{(2)}_{B}(0)$.  Using equation \ref{ord2curAB} and \ref{jc1B} we find

\begin{widetext}
\begin{align}
\bm{j}^{(2)}_{C1B}(0)+\bm{j}^{(2)}_{B}(0)=\sum_{\bm{k},n,m,i,j,\omega'=\pm \omega}\dfrac{e^3}{V\hbar}\dfrac{(f_n^T(\bm{k},\mu)-f_m^T(\bm{k},\mu))v_{nm}^i(\bm{k})}{\varepsilon_n(\bm{k})-\varepsilon_m(\bm{k})+\hbar\omega'+i\hbar\eta}
( \bm{\nabla}_{\bm{k}}v_{mn}^j(\bm{k})+i v_{mn}^j(\bm{k})(\bm{R}_{nn}(\bm{k})-\bm{R}_{mm}(\bm{k})))\nonumber\\A^i(\omega')A^j(-\omega')
\label{sumcur}
\end{align}
\end{widetext}

\noindent
We can now break up the denominator into its resonant and principle parts leading to two contributions to $\bm{j}^{(2)}_{C1B}(0)+\bm{j}^{(2)}_{B}(0)$.

We now will combine $\bm{j}^{(2)}_{C1A}(0)$, $\bm{j}^{(2)}_{C1B}(0)$, and $\bm{j}^{(2)}_{B}(0)$.  We use the identity 

\begin{equation}
-\mathcal{P}\dfrac{\bm{v}_{nn}(\bm{k})-\bm{v}_{mm}(\bm{k})}{2(\varepsilon_n(\bm{k})-\varepsilon_m(\bm{k})+\hbar\omega')^2}=\bm{\nabla}_{\bm{k}}\mathcal{P}\dfrac{1}{2\hbar(\varepsilon_n(\bm{k})-\varepsilon_m(\bm{k})+\hbar\omega')}
\end{equation}

\noindent
in the principle part of $\bm{j}^{(2)}_{C1A}(0)$ and then integrate by parts.  This leads to a term proportional to $\bm{k}$-space derivatives of the Fermi occupation factors and a contribution equal to the negate of the principle part of equation \ref{sumcur}.  We note that terms in equation \ref{sumcur} proportional to matrix elements of the dipole operator have no principle contribution as can be shown by the relabeling of dummy indices ($i\leftrightarrow j$, $\omega \leftrightarrow -\omega$, and $n \leftrightarrow m$).  

This leads to the following three contributions to the current arising from the $\bm{j}^{(2)}_{C1A}(0)$, $\bm{j}^{(2)}_{C1B}(0)$, and $\bm{j}^{(2)}_{B}(0)$ contributions to $\bm{j}^{(2)}(0)$.  We can write them as

\begin{widetext}
\begin{gather}
\bm{j}^{\rm inj}(0)=-\sum_{\bm{k},n,m,i,j,\omega'=\pm \omega}\dfrac{\pi e^3}{2V\hbar\eta} (f_n^T(\bm{k},\mu)-f_m^T(\bm{k},\mu))(\bm{v}_{nn}(\bm{k})-\bm{v}_{mm}(\bm{k}))v^i_{nm}(\bm{k})v^j_{mn}(\bm{k})\delta(\varepsilon_n(\bm{k})-\varepsilon_m(\bm{k})+\hbar\omega')\nonumber\\ \times A^i(\omega' )A^j(-\omega')\nonumber \\
\bm{j}^{\rm shift}(0)=- \sum_{\bm{k},n,m,i,j,\omega'=\pm \omega}\dfrac{e^3\pi}{V\hbar} (f_n^T(\bm{k},\mu)-f_m^T(\bm{k},\mu)) (v_{nm}^i(\bm{k})i\bm{\nabla}_{\bm{k}}v^j_{mn}(\bm{k})+v_{nm}^i(\bm{k})v_{mn}^j(\bm{k})(\bm{R}_{mm}(\bm{k})-\bm{R}_{nn}(\bm{k}))) \nonumber \\
\times \delta(\varepsilon_n(\bm{k})-\varepsilon_m(\bm{k})+\hbar\omega') A^i(\omega')A^j(-\omega') \nonumber \\
\bm{j}^{\rm Fermi}(0)= -\sum_{\bm{k},n,m,i,j,\omega'=\pm \omega}\dfrac{e^3}{V\hbar}\bm{\nabla}_{\bm{k}}f_n^T(\bm{k},\mu) \dfrac{v_{nm}^i(\bm{k})v_{mn}^j(\bm{k})}{\varepsilon_n(\bm{k})-\varepsilon_m(\bm{k})+\hbar\omega'}A^i(\omega')A^j(-\omega').
\label{j2con}
\end{gather}
\end{widetext}

The first term, $\bm{j}^{\rm inj}(0)$, is the injection current.  It is divergent for $\eta\rightarrow0$, but can be regulated with a relaxation time $\tau(\bm{k})$ which amounts to adding a term in the equation of motion for the density matrix (equation \ref{vonE}) of the form

\begin{equation}
\dfrac{\rho_{nm}(\bm{k},\omega)-\delta_{\omega,0}\delta_{nm}f_n^T(\bm{k},\mu)}{\tau_{nm}(\bm{k})}
\end{equation}

\noindent
This term encodes the effects of other interactions in the Hamiltonian not considered above that are linear in the deviations of the density matrix from its equilibrium value, $\rho_{nm}^{(0)}(\bm{k},\omega)$.  Its inclusion into the equation of motion of the density matrix ultimately amounts to substitution of $\eta$ for $1/\tau_{nm}(\bm{k})$.  The injection current, $\bm{j}^{\rm inj}(0)$, can be shown to vanish for time reversal symmetric systems perturbed by linear polarized light, by noting the constraint time reversal symmetry puts on the matrix elements of the velocity operator: $\bm{v}_{nm}(\bm{k})=-\bm{v}_{mn}(-\bm{k})$.

The second term in equation \ref{j2con}, $\bm{j}^{\rm shift}(0)$, is the shift current.  It is an intrinsic contribution to $\bm{j}^{(2)}(0)$ arising from $\bm{k}$-space derivatives of matrix elements of the velocity matrix.  These $\bm{k}$-space derivatives are accompanied with diagonal elements of the dipole operator such that the current remains invariant under a gauge transformation of the Bloch states of the form $\ket{u_n(\bm{k})}\rightarrow e^{i\phi_n(\bm{k})} \ket{u_n(\bm{k})}$ for all $n$.  It can be shown to vanish for time reversal symmetric materials exposed to external electric fields with circular polarization, but is nonzero for electric fields with linear polarization.

Both the injection and shift current derive from electronic transition between Bloch electrons with energy differences $\varepsilon_n(\bm{k})-\varepsilon_m(\bm{k})=-\hbar\omega'$ independent of crystal momentum $\bm{k}$.  Connection with the canonical expressions for these currents in { \it length} gauge can be made using this constraint upon replacement of the electromagnetic vector potential with the electric field and the use of the relationship between off diagonal elements of the velocity and dipole operators.

Lastly the third term in equation \ref{j2con}, $\bm{j}^{\rm Fermi}(0)$, is a contribution to the current proportional to $\bm{k}$-space derivatives of the equilibrium Fermi occupation factors $f_n^T(\bm{k},\mu)$.  At zero temperature $\bm{\nabla}_{\bm{k}}f_n^{T=0}(\bm{k},\mu)\rightarrow \hbar\bm{v}_{nn}(\bm{k}) \delta(\varepsilon_n(\bm{k})-\mu)$ such that the sum over the Brillouin zone is constrained to crystal momentum $\bm{k}$ along the Fermi surface.  For time reversal symmetric systems this contribution to the current can be shown to vanish for external electric fields with linear polarization, but can be nonzero for circular polarized light. %Below we calculate this contribution to $\bm{j}^{(2)}(0)$ on a simple model of a Weyl semimetal and show under certain conditions this contribution to the current is proportional to the charge of the Weyl point enclosed by the Fermi surface.

The last contributions to $\bm{j}^{(2)}(0)$ are from $\bm{j}^{(2)}_A(0)$ and $\bm{j}^{(2)}_{C0}(0)$.  From equation \ref{vonN2} we find 

\begin{align}
\bm{j}^{(2)}_{C0}(0)&=\sum_{\bm{k},n,m,i,j,\omega'=\pm \omega}\dfrac{e^3}{2V\hbar^2}\dfrac{f_n(\bm{k})-f_m(\bm{k})}{\varepsilon_n(\bm{k})-\varepsilon_m(\bm{k})}\bm{v}_{mn}(\bm{k}) \nonumber\\
&\times\bra{u_n(\bm{k})}\partial_{k_i}\partial_{k_j}\hat{H}_0(\bm{k})\ket{u_m(\bm{k})}A_i(\omega')A_j(-\omega')
\label{jC0}
\end{align}

\noindent
We can perform an integration by parts of $\bm{j}_A^{(2)}(0)$ and add the expression to $\bm{j}^{(2)}_{C0}(0)$.  Using the relationship between matrix elements of the velocity operator and matrix elements of the dipole operator and the identity $\bra{u_n(\bm{k})}\partial_{k_i}\ket{u_m(\bm{k})}=-(\partial_{k_i}\bra{u_n(\bm{k})})\ket{u_m(\bm{k})}$ we find the final contribution to the current $\bm{j}^{\rm Fermi2}(0)=\bm{j}_A^{(2)}(0)+\bm{j}^{(2)}_{C0}(0)$ can be written as

\begin{align}
\bm{j}^{\rm Fermi2}(0)&=-\sum_{\bm{k},n,,i,j,\omega'=\pm \omega}\dfrac{e^3}{2V\hbar^3}\bra{u_n(\bm{k})}\partial_{k_i}\partial_{k_j}\hat{H}_0(\bm{k})\ket{u_n(\bm{k})} \nonumber \\
&\times\bm{\nabla}_{\bm{k}}f_n^T(\bm{k},\mu)A_i(\omega')A_j(-\omega')
\label{jFer2}
\end{align}

\noindent
At zero temperature, like $\bm{j}^{\rm Fermi2}(0)$, this term $\bm{j}^{\rm Fermi2}(0)$ has $\bm{k}$-space contributions constrained to crystal momentum along the Fermi surface.
Due to the second-order derivative of Hamiltonian in $\bm{k}$ space, $\bm{j}^{\rm Fermi2}(0)$ is nonzero for systems whose energy dispersion relation $H(\bm{k})$ is higher than second order in $\bm{k}$. As shown in Eq. \ref{jFer2}, $\bm{j}^{\rm Fermi2}(0)$ is independent of TRS but only depends upon the polarization of the light. Since CP light is antisymmetric ($A_i(\omega')A_j(-\omega') = -A_i(-\omega')A_j(\omega')$) while LP light is symmetric ($A_i(\omega')A_j(-\omega') = A_i(-\omega')A_j(\omega')$), $\bm{j}^{\rm Fermi2}(0)$ is nonvanishing only with LP light.
It is also straightforward to see $\bm{j}^{\rm Fermi2}(0)$ is gauge invariant.  Our derivation is consistent with formulas derived using Feynman diagrammatic approach\cite{parker2019diagrammatic,matsyshyn2020berry}.

\section{Equivalence between velocity gauge and length gauge}
The equivalence between velocity gauge and length gauge can be proved if we rewrite our $\bm{j}^{\rm Fermi}(0)$ (Eq. \ref{j2con}) and $\bm{j}^{\rm Fermi2}(0)$ (Eq. \ref{jFer2}), which is composed of $\bm{j}^{(2)}_A(0)$ (Eq. \ref{jA2})\ and $ \bm{j}^{(2)}_{C0}(0)$ (Eq. \ref{jC0}). According to the useful relation of a general operator $\hat{O}(\bm{k})$~\cite{ventura2017gauge},
\begin{align}
    \bra{u_n(\bm{k})}\bm{\nabla} \hat{O}(\bm{k})\ket{u_{m}(\bm{k})} = \nabla\bra{u_n(\bm{k})}\hat{O}(\bm{k})\ket{u_{m}(\bm{k})} - i\bra{u_n(\bm{k})}[\bm{R}(\bm{k}),\hat{O}(\bm{k})]\ket{u_{m}(\bm{k})},
    \label{relation}
\end{align}
$\bm{j}^{(2)}_A(0)$ can be rewritten as:
\begin{gather}
    \bm{j}^{(2)}_{A}(0) = \sum_{\bm{k},n,m,i,j,\omega'=\pm\omega}\frac{e^3}{2V\hbar^3}\bra{u_n(\bm{k})}\bm{\nabla}\partial_{k_{i}} \partial_{ k_{j}}\hat{H}_{0}(\bm{k})\ket{u_{m}(\bm{k})}f_n(\bm{k})A_i(\omega')A_j(-\omega')\nonumber  \\
    =\sum_{\bm{k},n,m,i, j,\omega'=\pm\omega}\frac{e^3}{2V\hbar^3}\big(\bm{\nabla}\bra{u_n(\bm{k})}\partial_{ k_{i}} \partial_{k_{j}}\hat{H}_{0}(\bm{k})\ket{u_{n}(\bm{k})}-i\bra{u_n(\bm{k})}[\bm{R}(\bm{k}),\partial_{k_{i}} \partial_{ k_{j}}\hat{H}_{0}(\bm{k})]\ket{u_n(\bm{k}}\big)f_n(\bm{k})A_i(\omega')A_j(-\omega'). 
    \label{jA}
\end{gather}
We denote the first term in Eq. \ref{jA} as $\bm{j}^{(2)}_{A1}(0)$, and further simplify it as: 
\begin{align}
    \bm{j}^{(2)}_{A1}(0) &= \sum_{\bm{k},n,m,i,j,\omega'=\pm\omega}-\frac{e^3}{2V\hbar^3}\bra{u_n(\bm{k})}\partial_{k_{i}} \partial_{k_{j}}\hat{H}_{0}(\bm{k})\ket{u_{n}(\bm{k})}\nabla_{\bm{k}}f_{n}(\bm{k})A_i(\omega')A_j(-\omega') \nonumber \\ 
    &= \sum_{\bm{k},n,m,i, j,\omega'=\pm\omega}-\frac{e^3}{2V\hbar^2}\big(\partial_{k_{i}}v_{nn}^{j}(\bm{k})-i\frac{\bra{u_n(\bm{k})}[R_i,\partial_{ k_{j}}\hat{H}_{0}(\bm{k})]\ket{u_n(\bm{k}}}{\hbar}\big)\nabla_{\bm{k}}f_{n}(\bm{k})A_i(\omega')A_j(-\omega') \nonumber
    \\
    &=\sum_{\bm{k},n, m, i,j,\omega'=\pm\omega} \frac{e^3}{2V\hbar^2} \big(v_{nn}^{j}(\bm{k})\partial_{k_{i}}\nabla_{\bm{k}} f_{n}(\bm{k})  + iR^{i}_{nm}(\bm{k}) v^{j}_{mn}(\bm{k})\nabla_{\bm{k}} f_{nm}(\bm{k})\big)A_i(\omega')A_j(-\omega').
    \label{jA1}
\end{align}
 Here and below $\varepsilon_{nm}(\bm{k}) = \varepsilon_{n}(\bm{k}) - \varepsilon_{m}(\bm{k})$ and $f_{nm}(\bm{k}) = f_{n}(\bm{k}) - f_{m}(\bm{k})$ are defined as the band energy and band occupation difference, respectively. The last step is obtained by switching band indices $n \leftrightarrow m$.  In Eq. \ref{jA1}, the first term proportional to $\partial_{k_{i}}\nabla_{\bm{k}} f_{n}(\bm{k})$ is referred to as the classical Drude contribution derived in length gauge~\cite{watanabe2021chiral}, which describes the occupation change perturbed by the electric field at second order. This can be captured by semiclassical transport theory even in the single band case.
 
Using Eq. \ref{vtor}, by converting $\bm{v}_{mn}(\bm{k})$ to $\bm{R}_{mn}(\bm{k})$, the second term in $\bm{j}^{(2)}_{A1}(0)$ is written as:
\begin{align}
     \bm{j}^{(2)}_{A1-b}(0) = \sum_{\bm{k},n,m,i,j,\omega'=\pm\omega}-\frac{e^3}{2V\hbar^3}\varepsilon_{mn}(\bm{k})R_{nm}^{i}(\bm{k})R_{mn}^{j}(\bm{k})\nabla_{\bm{k}}f_{nm}(\bm{k})A_i(\omega')A_j(-\omega')\nonumber\\
     = \sum_{\bm{k},n,m,i,j,\omega'=\pm\omega}\frac{e^3}{2V\hbar^3}\big(\hbar\omega' + (\varepsilon_{nm}(\bm{k})- \hbar \omega' )\big)R_{nm}^{i}(\bm{k})R_{mn}^{j}(\bm{k})\nabla_{\bm{k}}f_{nm}(\bm{k})A_i(\omega')A_j(-\omega').
     \label{jA1-2}
\end{align}
 In Eq. \ref{jA1-2}, we break $\bm{j}^{(2)}_{A1-b}(0)$ into a piece with photon energy $\hbar \omega'$. This is the Berry curvature dipole contribution derived in length gauge~\cite{sodemann2015quantum, de2020difference,watanabe2021chiral}. To prove that, we rewrite this term as:
 \begin{align}
     \bm{j}^{(2)}_{A1-b1}(0) = \sum_{\bm{k},n,m,i,j,\omega'=\pm\omega}\frac{e^3\hbar\omega' }{2V\hbar^3}\big(R_{nm}^{i}(\bm{k})R_{mn}^{j}(\bm{k})-R_{nm}^{j}(\bm{k})R_{mn}^{i}(\bm{k})\big)\nabla_{\bm{k}}f_{n}(\bm{k})A_i(\omega')A_j(-\omega').
     \label{JA1-b1}
 \end{align}
 Using Eq. 13 in \cite{aversa1995nonlinear}:
 \begin{align}
     \frac{\partial R_{nn}^{j}(\bm{k})}{\partial k_i} - \frac{\partial R_{nn}^{i}(\bm{k})}{\partial k_j} = i\sum_{m} \big( R_{nm}^{i}(\bm{k})R_{mn}^{j}(\bm{k})-R_{nm}^{j}(\bm{k})R_{mn}^{i}(\bm{k}) \big),
 \end{align}
 and the definition of Berry curvature $\bm{\Omega}(\bm{k})$:
 \begin{align}
     \bm{\Omega}_{n}(\bm{k}) = \bm{\nabla}_{\bm{k}} \cross \bm{R}_{nn}(\bm{k}), 
 \end{align}
 Eq. \ref{JA1-b1} can be rewritten as:
 \begin{align}
     \bm{j}^{(2)}_{A1-b1}(0) = \sum_{\bm{k},n,i,j,\omega'=\pm\omega}-\frac{ie^3\hbar\omega' }{2V\hbar^3}\epsilon^{ijl}\Omega^{l}_n(\bm{k})\bm{\nabla_{k}}f_{n}(\bm{k})A_i(\omega')A_j(-\omega'),
 \end{align}
 which is the same as three-index tensor derived in length gauge \cite{de2020difference}.
 The remaining term in $\bm{j}^{(2)}_{A1-b}(0)$ (we denote it as $\bm{j}^{(2)}_{A1-b2}(0)$) can be added to $\bm{j}^{\rm Fermi}(0)$. $\bm{j}^{\rm Fermi}(0)$ is equivalent to:
\begin{align}
    \bm{j}^{\rm Fermi}(0) = \sum_{\bm{k},n,m,i, j,\omega'=\pm\omega} -\frac{e^3}{2V\hbar} \frac{v_{nm}^{i}(\bm{k})v_{mn}^{j}(\bm{k})}{\varepsilon_{nm}(\bm{k}) + \hbar \omega'} \nabla_{\bm{k}}f_{nm}(\bm{k})A_i(\omega')A_j(-\omega') \nonumber
    \\= \sum_{\bm{k},n,m,i, j,\omega'=\pm\omega} \frac{e^3}{2V\hbar^3} \frac{\varepsilon_{nm}(\bm{k}) \varepsilon_{mn}(\bm{k})R_{nm}^{i}(\bm{k})R_{mn}^{j}(\bm{k})}{\varepsilon_{nm}(\bm{k}) + \hbar \omega'} \nabla_{\bm{k}} f_{nm}(\bm{k})A_i(\omega')A_j(-\omega').
\end{align}
We thus arrive:
\begin{align}
    \bm{j}^{(2)}_{A1-b2}(0) + \bm{j}^{\rm Fermi}(0) = \sum_{\bm{k},n,m,i,j,\omega'=\pm\omega}-\frac{e^3}{2V\hbar^3}\frac{(\hbar \omega')^{2}R_{nm}^{i}(\bm{k})R_{mn}^{j}(\bm{k})}{\varepsilon_{nm}(\bm{k}) + \hbar \omega'}\nabla_{\bm{k}} f_{nm}(\bm{k})A_i(\omega')A_j(-\omega'),
    \label{jFermi-length}
\end{align}
which is the $\bm{j}^{\rm Fermi}$ derived in length gauge~\cite{de2020difference}. We note the difference between $\bm{j}^{\rm Fermi}$ derived in two gauges is $\bm{v}_{nm}(\bm{k})$ is replaced with $\bm{R}_{nm}(\bm{k})$.

Reviewing $\bm{j}^{(2)}_{A}(0)$(Eq. \ref{jA}), apart from $\bm{j}^{(2)}_{A1}(0)$ discussed above, the remaining term $\bm{j}^{(2)}_{A2}(0)$ is written as:
\begin{gather}
    \bm{j}^{(2)}_{A2}(0) = \sum_{\bm{k},n,m,i,j,\omega'=\pm\omega}-\frac{e^3 i}{2V\hbar^3} \big( \bm{R}_{nm}(\bm{k})\bra{u_m(\bm{k})}\partial_{k_{i}} \partial_{ k_{j}}\hat{H}_{0}(\bm{k})\ket{u_{n}(\bm{k})} - \bra{u_n(\bm{k})}\partial_{k_{i}} \partial_{k_{j}}\hat{H}_{0}(\bm{k})\ket{u_{m}(\bm{k})}\bm{R}_{mn}(\bm{k})\big)\nonumber \\ \times f_{n}(\bm{k})A_i(\omega')A_j(-\omega')\nonumber\\
    = \sum_{\bm{k},n,m,i,j,\omega'=\pm\omega}-\frac{e^3 i}{2V\hbar^2}\bm{R}_{nm}(\bm{k})\big(\partial_{k_i}v^{j}_{mn}(\bm{k})-i\bra{u_m(\bm{k})}[R_i, v^{j}(\bm{k})]\ket{u_n(\bm{k}}\big)f_{nm}(\bm{k})A_i(\omega')A_j(-\omega') \nonumber \\
    = \sum_{\bm{k},n,m,i,j,\omega'=\pm\omega}-\frac{e^3 i}{2V\hbar^2}\bm{R}_{nm}(\bm{k})\Big(\partial_{k_i}v^{j}_{mn}(\bm{k})-i\sum_{l}\big(R_{ml}^{i}(\bm{k})v^{j}_{ln}(\bm{k}) - v_{ml}^{j}(\bm{k})R^{i}_{ln}(\bm{k})\big) \Big)f_{nm}(\bm{k})A_i(\omega')A_j(-\omega').
\end{gather}

 Similarly, $\bm{j}^{(2)}_{C0}(0)$ (Eq. S70) is simplified as: 
\begin{gather}
    \bm{j}^{(2)}_{C0}(0)= \sum_{\bm{k},n,m,i,j,\omega'=\pm\omega}\dfrac{e^3}{2V\hbar^2}f_{nm}(\bm{k})\frac{\bm{v}_{mn}(\bm{k})}{\varepsilon_{nm}(\bm{k})}\bra{u_n(\bm{k})}\partial_{k_{i}} \partial_{k_{j}}\hat{H}_{0}(\bm{k})\ket{u_{m}(\bm{k})}A_i(\omega')A_j(-\omega') \nonumber \\
    =\sum_{\bm{k},n,m,i,j,\omega'=\pm\omega}-\frac{e^3 i}{2V\hbar^2}\bm{R}_{mn}(\bm{k})
    \Big(\partial_{k_i}v^{j}_{nm}(\bm{k})-i\sum_{l}\big(R_{nl}^{i}(\bm{k})v^{j}_{lm}(\bm{k}) - v_{nl}^{j}(\bm{k})R^{i}_{lm}(\bm{k})\big) \Big)f_{nm}(\bm{k})A_i(\omega')A_j(-\omega').
    \label{jc0-simple}
\end{gather}
By switching $n \leftrightarrow m$, $ \bm{j}^{(2)}_{C0}(0)$ is cancelled with $\bm{j}^{(2)}_{A2}(0)$.

As a summary, the sum of $\bm{j}^{\rm Fermi}(0)$ and $\bm{j}^{\rm Fermi2}(0)$ derived in velocity gauge (Eqs. \ref{j2con} and \ref{jFer2}, respectively) are equivalent to the sum of three contributions derived in length gauge: the Drude contribution ($\bm{j}^{\rm Drude}(0)$), Berry curvature dipole contribution ($\bm{j}^{\rm BCD}(0)$), and Fermi surface contribution in length gauge ($\bm{j}^{\rm Fermi-length}(0)$): 
\begin{align}
    \bm{j}^{\rm Drude}(0)=\sum_{\bm{k},n, i, j,\omega'=\pm\omega} \frac{e^3}{2V\hbar^2} v_{nn}^{j}(\bm{k})\partial_{k_{i}}\nabla_{\bm{k}} f_{n}(\bm{k})A_i(\omega')A_j(-\omega'),
\end{align}

 \begin{align}
     \bm{j}^{\rm BCD}(0) = \sum_{\bm{k},n,m,i,j,\omega'=\pm\omega}-\frac{ie^3\hbar\omega' }{2V\hbar^3}\epsilon^{ijl}\Omega^{l}_n(\bm{k})\bm{\nabla_{k}}f_{n}(\bm{k})A_i(\omega')A_j(-\omega'),
 \end{align}

 \begin{align}
    \bm{j}^{\rm Fermi-length}(0) = \sum_{\bm{k},n,m,i,j,\omega'=\pm\omega}-\frac{e^3}{2V\hbar^3}\frac{(\hbar \omega')^{2}R_{nm}^{i}(\bm{k})R_{mn}^{j}(\bm{k})}{\varepsilon_{nm}(\bm{k}) + \hbar \omega'}\nabla_{\bm{k}} f_{nm}(\bm{k})A_i(\omega')A_j(-\omega').
 \end{align}

\section{DC Fermi Surface Currents in Weyl Semimetals}

 We consider the incident CP light with a vector field $\boldsymbol{A}(t)=A_{0}(\cos{\omega t}\vec{e}_{x}+\sin{\omega t}\vec{e}_y)$. Using band-resolved Berry curvature $\Omega^{z}_{nm}(\boldsymbol{k})=i(R_{nm}^{x}R_{mn}^{y}-R_{nm}^{y}R_{mn}^{x})$, when $T\rightarrow 0$, $\boldsymbol{j}^{Fermi}$ is:
 \begin{widetext}
 \begin{align}
     {j}^{\rm Fermi,z}= \frac{e^3 i }{V\hbar^{2}}\sum_{n,m,\boldsymbol{k}}\frac{2\hbar\omega\boldsymbol(\varepsilon_{n}(\boldsymbol{k})-\varepsilon_{m}(\boldsymbol{k}))^2}{(\varepsilon_{n}(\boldsymbol{k})-\varepsilon_{m}(\boldsymbol{k}))^{2}-\hbar^{2}\omega^{2}}\delta(\varepsilon_{n}(\boldsymbol{k})-\mu){v}^{z}_{nn}(\boldsymbol{k})\Omega^{z}_{nm}(\boldsymbol{k})A^{x}(\omega)A^{y}(-\omega).
 \end{align}
 \end{widetext}
 For isotropic Weyl semimetal with two bands, the current is:
  \begin{widetext}
 \begin{align}
     {j}^{\rm Fermi,z}= \frac{e^3 i }{V\hbar^{2}}\frac{2\omega\boldsymbol(2\varepsilon_{F})^{2}}{(2\varepsilon_{F})^{2}-\hbar^{2}\omega^{2}}\sum_{\boldsymbol{k}}\delta(k-k_F)\hat{n}^{z}(\boldsymbol{k})\Omega^{z}(\boldsymbol{k})A^{x}(\omega)A^{y}(-\omega),
 \end{align}
 \end{widetext}
 where $\varepsilon_F$ is the Fermi energy relative to the Weyl node, $k_F$ is the magnitude of the Fermi vector, $\hat{\boldsymbol{n}}(\boldsymbol{k})$ is the normal vector to the Fermi surface, and $\boldsymbol{\Omega({k})}=(\Omega_{12}^{x}{(\boldsymbol{k})},\Omega_{12}^{y}{(\boldsymbol{k})},\Omega_{12}^{z}{(\boldsymbol{k})})$, where $1,2$ refer to the low and high energy bands, respectively. For a CP light, the current can also be defined through $j^{i} = \chi_{ij}(\omega)[\boldsymbol{E}(\omega)\times\boldsymbol{E}^{*}(\omega)]_{j}$, where $\chi_{ij}$ is an imaginary tensor. For CP light adopted above, $[\boldsymbol{E}(\omega)\times\boldsymbol{E}^{*}(\omega)] \vec{e_z} =2E_{x}(\omega)E_y(-\omega)=2\omega^2 A_{x}(\omega)A_y(-\omega) \vec{e_z}= iE_{0}^{2}/2\ \vec{e_z}=i\omega^{2}A_{0}^{2}/2\ \vec{e_z}$, and ${j}^{Fermi,z} = \chi_{zz}[\boldsymbol{E}(\omega)\times\boldsymbol{E}^{*}(-\omega)]_{z}$. ${j}^{Fermi,x}$ and ${j}^{Fermi,y}$ can be established in an analogous approach, and the total $\boldsymbol{j}^{Fermi}$ is related with the trace over $\chi_{ij}$:
 \begin{align}
     \text{Tr}[\chi_{ij}]&= \frac{e^3 i }{2V\hbar^{2}}\frac{\boldsymbol(2\varepsilon_{F})^{2}}{\omega[(2\varepsilon_{F})^{2}-\hbar^{2}\omega^{2}]}\sum_{\boldsymbol{k}}\delta(k-k_F)\hat{\boldsymbol{n}}(\boldsymbol{k})\cdot\boldsymbol{\Omega}(\boldsymbol{k})\nonumber\\
     &=\frac{e^3 i }{2\hbar^{2}}\frac{2\boldsymbol(2\varepsilon_{F})^{2}}{\omega[(2\varepsilon_{F})^{2}-\hbar^{2}\omega^2]}\int \frac{d\boldsymbol{k}}{(2\pi)^3}\delta(k-k_F)\hat{\boldsymbol{n}}(\boldsymbol{k})\cdot\boldsymbol{\Omega}(\boldsymbol{k})\nonumber\\
     &=i\frac{e^3 }{h^{2}}\frac{(2\varepsilon_{F})^{2}}{\omega[(2\varepsilon_{F})^{2}-\hbar^{2}\omega^{2}]}Q_{i},
 \end{align}
 where $Q_{n}=\frac{1}{2\pi}\int_{FS} d\boldsymbol{S}\cdot\boldsymbol{\Omega}(\boldsymbol{k})$ is the charge of the Weyl point $n$. For time-reversal invariant system, weyl points at $\boldsymbol{K}$ and $-\boldsymbol{K}$ have the same charge so won't annihilate. The overall contribution is: 
 \begin{align}
     \text{Tr}[\beta_{ij}]= i\frac{e^3 }{h^2}\sum_{n}\frac{(2\varepsilon_{Fn})^{2}}{\omega[(2\varepsilon_{Fn})^{2}-\hbar^{2}\omega^{2}]}Q_{i},
\end{align}
where $\varepsilon_{Fn}$ is the Fermi level relative to node $n$, and the sum is over all Weyl points in the first Brillouin Zone.

\bibliography{ref.bib}